\documentclass{aa}  

\usepackage{graphicx}
\usepackage{txfonts}
\usepackage{natbib}
\bibpunct{(}{)}{;}{a}{}{,}
\def\XMM{{\sl XMM-Newton}}
\def\swift{{\sl Swift}}
\def\chan{{\sl Chandra}}
\def\beppo{{\sl Beppo}-SAX}

\begin{document}

   \title{Multi-epoch properties of the warm absorber in the Seyfert 1 galaxy NGC 985}


   \author{J. Ebrero\inst{1}
          \and
          V. Dom\v{c}ek\inst{2,3,4}
          \and
          G. A. Kriss\inst{5}
          \and
          J. S. Kaastra\inst{6,7}
          }

   \institute{Telespazio UK for the European Space Agency (ESA),
     European Space Astronomy Centre (ESAC),
              Camino Bajo del Castillo, s/n, 
              E-28692 Villanueva de la Ca\~nada, Madrid, Spain\\
              \email{jebrero@sciops.esa.int}
              \and
              Anton Pannekoek Institute for Astronomy, University of Amsterdam, Science Park 904, 1098 XH Amsterdam, The Netherlands
              \and
              GRAPPA, University of Amsterdam, Science Park 904, 1098 XH Amsterdam, The Netherlands
              \and
              Department of Theoretical Physics and Astrophysics, Masaryk University, Kotl{\'a}{\v r}sk{\'a} 2, 61137, Czech Republic
              \and
              Space Telescope Science Institute, 3700 San Martin Drive, Baltimore, MD 21218, USA
              \and
              SRON Netherlands Institute for Space Research, Sorbonnelaan 2, 3584 CA Utrecht, The Netherlands
              \and
              Leiden Observatory, Leiden University, P. O. Box 9513, 2300 RA Leiden, The Netherlands
              }

   \date{Received <date>; accepted <date>}

 
  \abstract
   {NGC 985 was observed by \XMM{}~twice in 2015, revealing that the source was coming out from a soft X-ray obscuration event that took place in 2013. These kinds of events are possibly recurrent since a previous \XMM{}~archival observation in 2003 also showed signatures of partial obscuration.}
   {We have analyzed the high-resolution X-ray spectra of NGC 985 obtained by the Reflection Grating Spectrometer (RGS) onboard \XMM{}~in 2003, 2013, and 2015 in order to characterize the ionized absorbers superimposed to the continuum and to study their response as the ionizing flux varies.}
   {The spectra were analyzed with the SPEX fitting package and the photoionization code CLOUDY.}
   {We found that up to four warm absorber (WA) components were present in the grating spectra of NGC 985, plus a mildy ionized ($\log \xi \sim 0.2 - 0.5$) obscuring ($N_{\rm H} \sim 2 \times 10^{22}$~cm$^{-2}$) wind outflowing at $\sim -6\,000$~km s$^{-1}$. The absorbers have a column density that ranges from  $\sim 10^{21}$ to a few times $10^{22}$~cm$^{-2}$, and ionization parameters ranging from $\log \xi \sim 1.6$ to $\sim 2.9$. The most ionized component is also the fastest, moving away at $\sim -5\,100$~km s$^{-1}$, while the others outflow in two kinematic regimes, $\sim -600$~and $\sim -350$~km s$^{-1}$. These components showed variability at different time scales in response to changes in the ionizing continuum. Assuming that these changes are due to photoionization and recombination mechanisms, we have obtained upper and lower limits on the density of the gas. We used these limits to pinpoint the location of the warm absorbers, finding that the closest two components are at parsec-scale distances, while the rest may extend up to tens of parsecs from the central source. With these constraints on the density and location, we found that the fastest, most ionized WA component accounts for the bulk of the kinetic luminosity injected back into the interstellar medium (ISM) of the host galaxy, which is on the order of $0.8\%$~of the bolometric luminosity of NGC 985. According to the models, this amount of kinetic energy per unit time would be sufficient to account for cosmic feedback.}
   {Observations of the onset and conclusion of transient obscuring events in active galactic nuclei (AGN) are a key tool to understand both the dynamics and physics of the gas in their innermost regions, and also to study the response of the surrounding gas as the ionizing continuum varies.}


   \authorrunning{J. Ebrero et al.}
   \titlerunning{Multi-epoch properties of the warm absorber in NGC 985}

   \maketitle
%

\section{Introduction}
\label{intro}

Gravitational accretion of matter onto the supermassive black hole (SMBH) that resides in the center of most, if not all, galaxies is what powers active galactic nuclei (AGN, \citealt{Rees84}). This very efficient process produces large amounts of energy across the entire spectrum that can be detected at cosmological distances. Not only is radiation emitted, but also large amounts of matter are injected back into the interstellar medium (ISM) of the host galaxy. This process is known as feedback, and it is believed to play a major role in the coevolution of the SMBH and the galaxy (\citealt{DM05}; \citealt{HE10}).

More than half of the observed type-1 AGN present signatures of photoionized gas in their soft X-ray spectrum, the so-called warm absorbers (WA), that are seen as absorption features superimposed to the continuum (e.g., \citealt{Rey97}; \citealt{Geo98}; \citealt{Cre03}). With the advent of high-resolution X-ray spectroscopy after the launch of \XMM{}~and \chan{}, it was observed that these absorbers were blueshifted with respect to the rest frame of their host galaxy, typically by hundreds to thousands of km s$^{-1}$ (\citealt{Kaa00}; \citealt{Kas01}). Since they were outflowing, they were injecting kinetic energy and mass back into the ISM and therefore they were potential candidates to be a source of cosmic feedback. It is now considered that the regular WAs have, in general, little impact on their surroundings, with the exception of their most extreme version: outflows with very high velocity, ionization state, and column density, known as ultra-fast outflows (UFOs), which can have a significant impact on their host galaxy (\citealt{Tom12, Tom13}; \citealt{Laha16}).

The dominant launching mechanism of these outflows is still a matter of debate as they present a wide range of velocities, degrees of ionization, and column densities. They range from thermal winds arising the from inner edge of the torus, which is eroded by the intense radiation field (\citealt{KK01}), and winds launched from the accretion disk by radiative means (\citealt{PK04}) or magnetohydrodynamical mechanisms (\citealt{KK94}; \citealt{Fuk10a, Fuk10b}). In addition to this, after many years of observations and monitoring of AGN, another type of transient wind has been identified and studied: high-column density, fast, obscuring winds that eclipse the inner X-ray source of AGN on relatively short time scales. A number of objects have now shown this kind of transient phenomenon, for instance NGC 1365 (\citealt{Ris07}), Mrk 766 (\citealt{Ris11}), Mrk 335 (\citealt{Lon13, Lon19}), NGC 5548 (\citealt{Kaa14}), NGC 3783 (\citealt{Meh17}), and now also NGC 985 (\citealt{Ebr16}). 

NGC 985, also known as Mrk 1048, is a nearby ($z = 0.043$, \citealt{Fis95}), bright ($F_{\rm 0.3-10~keV} = 2.2 \times 10^{-11}$~erg cm$^{-2}$ s$^{-1}$, \citealt{Ebr16}) Seyfert 1 galaxy with a distinctive ring shape, produced by a recent galactic merger. This source is known to host a WA in X-rays (\citealt{Kro05,Kro09}) as well as an UV absorber (\citealt{Ara02}). During a \swift{}~monitoring program in 2013, the source was found to be in a very low soft X-ray flux state. A triggered joint \XMM{}~and HST observation found that some intervening very dense gas was occulting the central X-ray source (\citealt{Par14}). In the most recent observations in 2015, taken 12 days apart also with \XMM{}~and HST, we found that the source was emerging from this obscured state. The first analysis of the X-ray broad band and UV spectra were presented in \citet{Ebr16}. In this work we focus on the analysis of the high-resolution X-ray spectra taken by the RGS onboard \XMM{}~in 2015 and in the previous archival observations of this source.

This paper is organized as follows. In Sect.~\ref{data} we described the observations used in this work and the data reduction. In Sect.~\ref{seds} we construct and absorbed and unabsorbed spectral energy distributions (SED) that will be used in the data analysis in Sect.~\ref{analysis}. Our results are described and discussed in Sect.~\ref{results}~and \ref{discussion}, respectively. Finally, we summarize our findings in Sect.~\ref{conclusions}. We adopt a cosmological framework with $H_0 = 70$~km s$^{-1}$~Mpc$^{-1}$, $\Omega_M = 0.3$, and $\Omega_{\Lambda} = 0.7$. The quoted errors refer to 68.3\% confidence level unless otherwise stated. 


\section{Observations and data reduction}
\label{data}

In this paper we focus on the analysis of the high-resolution grating spectra of NGC 985 obtained by the \XMM{}~Reflection Grating spectrometer (RGS; \citealt{Her01}. We use the same observations as in \citet{Ebr16}, where the EPIC-pn data were analyzed: one observation in 2003 (ObsID 0150470601), one in 2013 (ObsID 0690870501), and two in 2015 (ObsIDs 0743830501 and 0743830601, respectively). Another observation in 2013 (ObsID 0690870101) was too short ($\sim$20~ks of exposure time) to provide an RGS spectrum with enough signal-to-noise, and therefore it was discarded from the present analysis. The observations were retrieved from the \XMM{}~Science Archive (XSA\footnote{\tt http://nxsa.esac.esa.int/nxsa-web/} and re-processed using SAS v16.0 (\citealt{Gab04}).

The RGS data were taken in the normal spectroscopy mode, and they were processed using {\it rgsproc} with cross-dispersion and CCD pulse-height selections of 95 per cent, and background exclusion of 98 per cent (the SAS default settings). The spectra were binned in beta space and the response matrix used 9\,000 bins. The observations and their exposure times are summarized in Table~\ref{xmmlog}.

   \begin{table}
     \centering
     \caption[]{\XMM{}~observations log of NGC 985.}
     \label{xmmlog}
     \begin{tabular}{l c c c}
       \hline\hline
       \noalign{\smallskip}
       ObsID & Obs. Date & Start Time & Exp. Time  \\
              &  (yyyy-mm-dd) & (hh:mm:ss)  & (ks)  \\
       \noalign{\smallskip}
       \hline
       \noalign{\smallskip}
       0150470601 & 2003-07-15 & 15:28:36 & 57.9 \\
       0690870501 & 2013-08-10 & 21:17:52 & 103.7 \\
       0743830501 & 2015-01-13 & 09:52:59 & 138.9 \\
       0743830601 & 2015-01-25 & 08:48:22 & 122.0 \\
       \noalign{\smallskip}
       \hline
     \end{tabular}
   \end{table}


\section{Obscured and unobscured spectral energy distributions}
\label{seds}

\begin{figure*}
  \centering
  \hbox{
  \includegraphics[width=6.5cm,angle=-90]{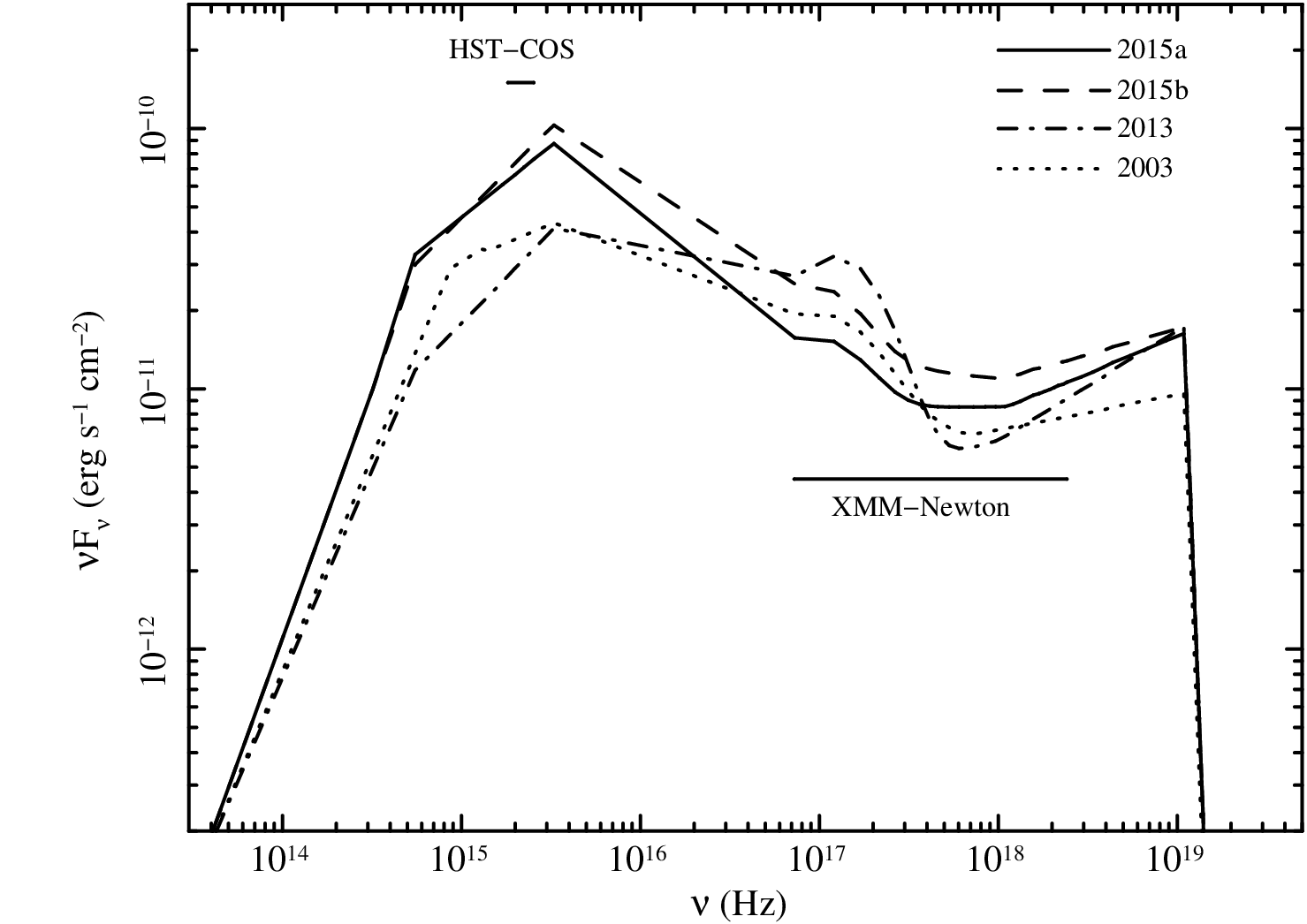}
  \includegraphics[width=6.5cm,angle=-90]{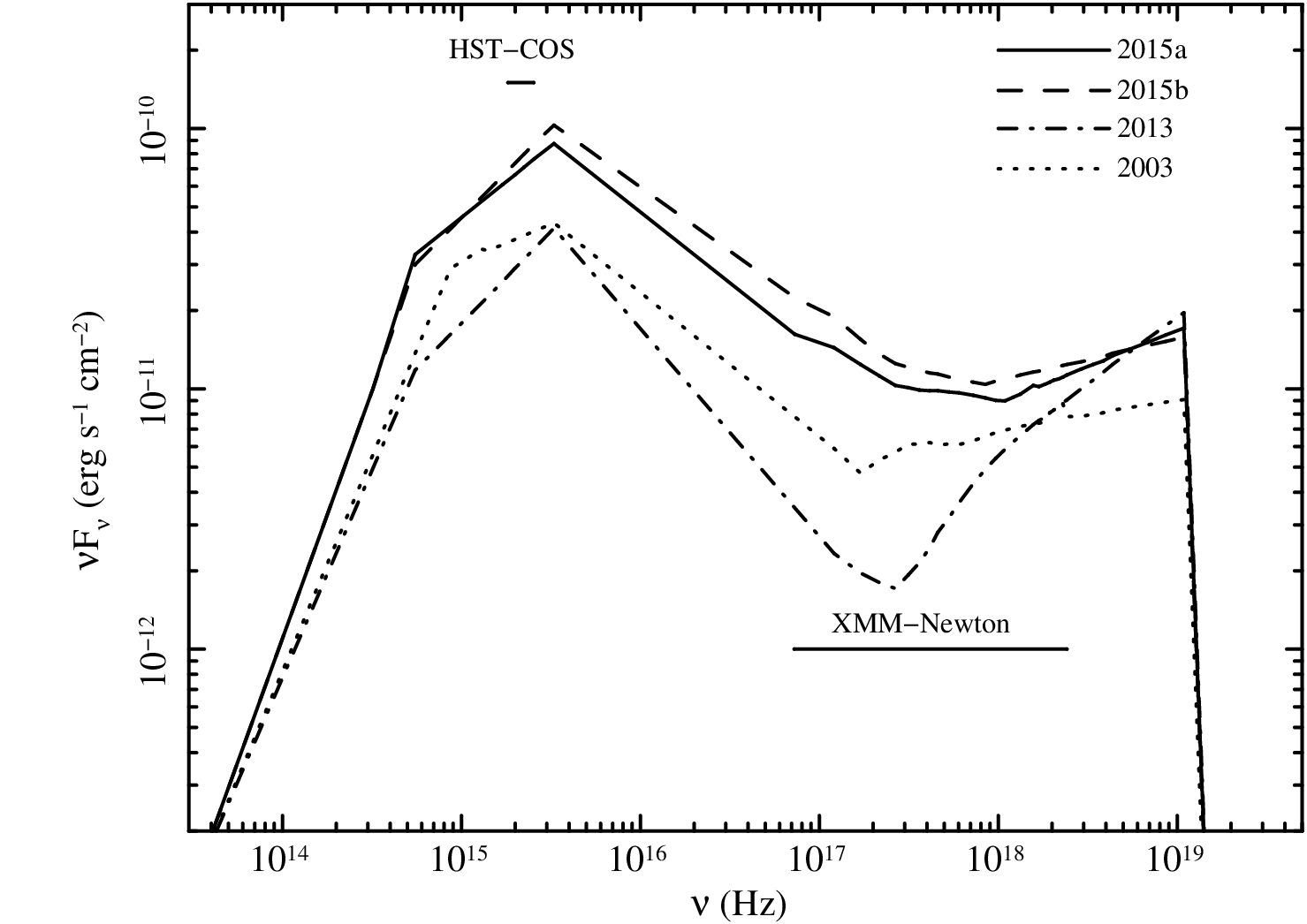}
  }
  \caption{\label{sedfig}Unobscured ({\it left panel}) and obscured ({\it right panel}) spectral energy distributions of the 2015a (solid line), 2015b (dashed line), 2013 (dot-dashed line), and 2003 (dotted line) observations. The bands covered by HST-COS and \XMM{}~are also indicated.}
\end{figure*}

The spectral energy distributions (SED) of the observations were constructed using the contemporaneous measurements of \XMM{}~EPIC-pn in the X-rays, HST in the UV (except for the 2003 observation in which no contemporaneous HST data were available), and the \XMM{}~optical monitor (OM) in the UV and optical band. Following the results of \citet{Ebr16}, where it is shown that the central ionizing source is partially blocked by obscuring material, we constructed two SED sets. One is the unobscured SED (e.g., the spectrum emitted by the AGN in NGC 985) and the other one is the obscured SED, filtered by the obscuring intervening material. The latter will be used in the analysis of the WA components in NGC 985, which are assumed to be located farther than the obscurer (see Sect.~\ref{analysis}). This approach was also followed in \citet{Meh15}~and \citet{Gesu15} in order to analyze the outflows in NGC 5548, which was also found to be in a persistent obscured state (\citealt{Kaa14}).

In both cases the X-ray SED was obtained from the best-fit EPIC-pn model (\citealt{Ebr16}) corrected from intrinsic and Galactic absorption. For the obscured SED we kept the obscuring component as measured in \citet{Ebr16}. In the UV we used the HST-COS continuum fluxes at 1175, 1339, 1510, and 1765~\AA~corrected for Galactic extinction using the reddening curve prescription of \citet{Car89}, assuming a color excess $E(B-V) = 0.03$~mag based on calculations of \citet{Sch98} as updated by \citet{Sch11}, as given in the NASA/IPAC Extragalactic Database (NED\footnote{\tt http://ned.ipac.caltech.edu/}), and a ratio of total to selective extinction $R_{\rm V} \equiv A_{\rm V}/E(B-V)$~fixed to 3.1.

The OM observations were taken with the $V$, $B$, $U$, $UVW1$, $UVM2$, and $UVW2$ filters in the 2013 and both 2015 observations, and with the $U$, $UVW1$, $UVM2$, and $UVW2$ filters in the 2003 observation. The OM fluxes were also corrected for Galactic extinction using the reddening curve of \citet{Car89}, including the update in the optical and near-IR of \citet{ODon94}. To correct for the host galaxy starlight contribution in the OM bandpasses we used the bulge galaxy template of \citet{Kin96}, scaled to the host galaxy flux at the rest-frame wavelength $F_{\rm gal,~5100~\AA} = 2.48 \times 10^{-15}$~erg cm$^{-2}$~s$^{-1}$~\AA$^{-1}$~(\citealt{Kim08}). The adopted SEDs are shown in Fig.~\ref{sedfig}.


\section{Data analysis}
\label{analysis}

The RGS spectra of NGC 985 were analyzed using the SPEX fitting package\footnote{\tt http://www.sron.nl/astrophysics-spex} version 3.05.01 (\citealt{Kaa96}). The adopted fitting method was C-statistics (\citealt{Cash79}) so that binning of the data is in principle no longer needed. However, to avoid oversampling, we rebinned by a factor of 5 in the $7 - 24$~\AA~range, and by a factor of 7 in the $24 - 38$~\AA~range, in order to compensate for lower effective area and therefore higher noise toward lower energies in the individual 2015 observations. The 2003 and 2013 observations, affected by a lower signal-to-noise due to the presence of the obscurer and shorter exposure time (in 2003), were analyzed only in the region $7 - 26$~\AA, rebbined by a factor of 7. The foreground Galactic column density was set to $N_{\rm H} = 3.17 \times 10^{20}$~cm$^{-2}$ (\citealt{Kal05}) using the {\it hot} model in SPEX, with the temperature fixed to 0.5~eV to mimic a neutral gas. Throughout the analysis we adopted a cosmological redshift of $z = 0.043$~(\citealt{Fis95}), and we assumed the proto-Solar abundances of \citet{LP09}.

\subsection{Continuum and emission lines}
\label{continuum}

The continuum was modeled locally in the RGS waveband with a single powerlaw  with the photon index $\Gamma$ fixed to the values obtained in the analysis of the EPIC-pn spectra of \citet{Ebr16}. The soft excess was modeled using a modified black body ({\it mbb} model in SPEX), which takes into account modifications of a simple black body model by coherent Compton scattering based on the calculations of \citet{KB89}. The best-fit values of the continuum parameters for the different observations are reported in Table~\ref{conttable}.

In addition, the RGS spectra showed the presence of some narrow emission lines sticking out of the continuum. These emission features were modeled with Gaussian line models with centroid wavelengths fixed to the laboratory wavelength at the rest-frame of the source. The most prominent line seen in all of the observations was the He-like forbidden \ion{O}{vii} line. In the 2013 observation, the depressed continuum due to the ongoing obscuration allowed to significally detect emission features from other species and transitions. H-like Ly$\alpha$ transitions of \ion{O}{viii} and \ion{Mg}{xii} were significantly detected in this observation, whereas that of \ion{N}{vii} was below the 2$\sigma$ level. The He-like forbidden line of \ion{Ne}{ix} was detected above the 4$\sigma$ level, whereas the He-like forbidden line of \ion{Mg}{xi} was less significant but still detected at the 3$\sigma$. The \ion{O}{vii} intercombination line was detected at the 2$\sigma$ level or higher in the 2003, 2013, and in the 2015a observation, while in the 2015b observation it was not significantly detected, owing to the higher continuum. We also detected the resonance line of He-like \ion{O}{vii}, above the 6$\sigma$ level, which was also found with less significance in the 2015b observation. Overall, the detection prominent forbidden lines together with slightly weaker resonance lines suggests that these lines originate in a photoionized plasma (\citealt{PD00}). The list of emission lines detected in the RGS spectra of NGC 985 together with their line fluxes and detection significance are reported in Table~\ref{emtable}.

In this Table we also show the laboratory wavelength and the measured best-fit, rest-frame centroid wavelengths for each transition. There is no evident connection between the emitting medium and the absorbers described in Sect.\ref{absorption}. The vast majority of the best-fit values are consistent with the laboratory ones, or slightly blueshifted with negligible significance (typically less than $1\sigma$). The gas responsible for the emission lines is therefore not ouflowing. Furthermore, the majority of the detected emission lines are narrow, unresolved lines. The most prominent of these lines, detected in all four observations, do not show changes in their widths. All of this implies that the emitting material is likely located much further away than the absorbers.

   \begin{table*}
     \centering
     \caption[]{Best-fit continuum parameters.}
     \label{conttable}
     \begin{tabular}{l|l c c c c}
       \hline\hline
       \noalign{\smallskip}
       Model & Parameter & Obs. 2003 & Obs. 2013 & Obs. 2015a & Obs. 2015b \\
       \noalign{\smallskip}
       \hline
       \noalign{\smallskip}
       Power-law & $F_{\rm 0.2-2~keV}$\tablefootmark{a} & $5.1^{+1.4}_{-0.8} \times 10^{-12}$ & $7.3^{+3.0}_{-2.1} \times 10^{-13}$ & $9.2\pm 1.7 \times 10^{-12}$ & $1.4 \pm 0.2 \times 10^{-11}$ \\
                 & $\Gamma$\tablefootmark{b}          & $2.21$ & $1.71$ & $2.14$ & $2.22$ \\
       Mod. BB   & $F_{\rm 0.2-2~keV}$\tablefootmark{a} & $1.4^{+3.1}_{-1.1} \times 10^{-13}$ & $7.5 \pm 0.4 \times 10^{-13}$ & $1.4 \pm 0.5 \times 10^{-12}$ & $2.3 \pm 0.5 \times 10^{-12}$  \\
                 & $T$\tablefootmark{c}               & $0.10 \pm 0.05$ & $0.27 \pm 0.04$ & $0.12 \pm 0.01$ & $0.12 \pm 0.01$ \\
       \noalign{\smallskip}
       \hline
     \end{tabular}
     \tablefoot{
       \tablefoottext{a}{Observed flux in the $0.2 - 2$~keV range, in units of erg cm$^{-2}$~s$^{-1}$;}
       \tablefoottext{b}{power-law photon index, frozen to the EPIC-pn best-fit values of \citet{Ebr16};}
       \tablefoottext{c}{temperature, in units of eV.}
     }
   \end{table*}

   \begin{table*}
     \centering
     \caption[]{Emission lines in the RGS spectra of NGC 985.}
     \label{emtable}
     \begin{tabular}{l c c c c c}
       \hline\hline
       \noalign{\smallskip}
       Line\tablefootmark{a} & $\lambda_0$\tablefootmark{b} & $\lambda$\tablefootmark{c} & $\Delta\lambda$\tablefootmark{d}  & Flux\tablefootmark{e} & $\sigma$\tablefootmark{f}  \\
       \noalign{\smallskip}
       \hline
       \noalign{\smallskip}
       \multicolumn{6}{c}{Obs. 2003} \\
       \noalign{\smallskip}
       \hline
       \noalign{\smallskip}
       \ion{O}{vii}, a$ w$ & 21.6015  & 21.3 $\pm$ 0.1 & 0.2 $\pm$ 0.1 & 28.0 $\pm$ 9.0  & 3.1 \\
       \ion{O}{vii}, a$ f$ & 22.0974 & 22.0 $\pm$ 0.1 & 0.4 $\pm$ 0.2 & 41.0 $\pm$ 10.0 & 4.1 \\
       \noalign{\smallskip}
       \hline
       \noalign{\smallskip}
       \multicolumn{6}{c}{Obs. 2013} \\
       \noalign{\smallskip}
       \hline
       \noalign{\smallskip}
       \ion{Mg}{xii}, L$\alpha$ & 8.421 & 8.41 $\pm$ 0.04 & $<$0.2 & 15.0 $\pm$ 5.0 & 3.0  \\ 
       \ion{Mg}{xi}, a$ f$ & 9.3136 & 9.25 $\pm$ 0.04 & 0.2 $\pm$ 0.1 & 15.0 $\pm$ 5.0 & 3.0 \\ 
       \ion{Ne}{ix}, a$ f$ & 13.6984 & 13.69 $\pm$ 0.02 & $<$0.1  & 6.4 $\pm$ 1.5 & 4.3 \\ 
       \ion{O}{viii}, L$\beta$ & 16.0059 & 16.1 $\pm$ 0.1 & 0.3 $\pm$ 0.1  & 8.4 $\pm$ 1.7 & 4.9 \\ 
       \ion{O}{vii}, b$ f$ & 18.7307 & 18.08 $\pm$ 0.02 & $<$0.2 & 6.2 $\pm$ 1.9 & 3.3 \\ 
       \ion{O}{viii}, L$\alpha$ & 18.9689 & 18.96 $\pm$ 0.02 & $<$0.2 & 12.0 $\pm$ 3.0 & 4.0 \\ 
       \ion{O}{vii}, a$ w$ & 21.6015 & 21.60 $\pm$ 0.04 & 0.3 $\pm$ 0.1 & 19.0 $\pm$ 3.0 & 6.3 \\ 
       \ion{O}{vii}, a$ i$ & 21.802 & 21.87 $\pm$ 0.03 & $<$0.1 & 6.2 $\pm$ 3.5 & 1.8 \\ 
       \ion{O}{vii}, a$ f$ & 22.0974 & 22.09 $\pm$ 0.01 & $<$0.1 & 21.0 $\pm$ 4.0 & 5.3 \\ 
       \ion{N}{vii}, L$\alpha$ & 24.781 & 24.7 $\pm$ 0.1 & 0.2$\pm$ 0.1 & 4.1 $\pm$ 2.6 & 1.6 \\ 
       \noalign{\smallskip}
       \hline
       \noalign{\smallskip}
       \multicolumn{6}{c}{Obs. 2015a} \\
       \noalign{\smallskip}
       \hline
       \noalign{\smallskip}
       \ion{O}{vii}, a$ i$ & 21.802 & 21.83 $\pm$ 0.03 & $<$0.1  & 11.0 $\pm$ 5.0 & 2.2 \\ 
       \ion{O}{vii}, a$ f$ & 22.0974 & 22.12 $\pm$ 0.02 & $<$0.1 & 21.0 $\pm$ 5.0 & 4.2 \\ 
       \ion{C}{vi}, L$\alpha$ & 33.735 & 33.6 $\pm$ 0.1 & $<$0.4 & 4.7 $\pm$ 2.9 & 1.6 \\ 
       \noalign{\smallskip}
       \hline
       \noalign{\smallskip}
       \multicolumn{6}{c}{Obs. 2015b} \\
       \noalign{\smallskip}
       \hline
       \noalign{\smallskip}
       \ion{O}{vii}, a$ w$ & 21.6015 & 21.57$\pm$ 0.05 & $<$0.4 & 23.0 $\pm$ 12.0 & 1.9 \\ 
       \ion{O}{vii}, a$ f$ & 22.0974 & 22.10$\pm$0.02 & $<$0.2 & 30.0 $\pm$ 11.0 & 2.7 \\
       \ion{C}{vi}, L$\alpha$ & 33.735 & 33.4 $\pm$ 0.1 & 0.3$\pm$0.1 & 35 $\pm$ 8.0 & 4.4 \\
       \noalign{\smallskip}
       \hline 
     \end{tabular}
     \tablefoot{
       \tablefoottext{a}{Emission line transitions (L: H-like, a: He$\alpha$-like, b: He$\beta$-like, $w$: resonance, $i$: intercombination, $f$: forbidden);}
       \tablefoottext{b}{laboratory wavelength, in \AA;}
       \tablefoottext{c}{rest-frame wavelength, in \AA;}
       \tablefoottext{d}{line width, in \AA;}
       \tablefoottext{e}{line flux, in units of $10^{-15}$~erg cm$^{-2}$~s$^{-1}$;}
       \tablefoottext{f}{significance of the detection.}
     }
   \end{table*}

\subsection{Absorption features}
\label{absorption}

NGC 985 is known to host an intrinsic WA in the form of ionized winds with multiple ionization phases (\citealt{Kro05}; \citealt{Kro09}). Indeed, the RGS spectra show a plethora of absorption troughs that can be attributed to intervening ionized gas at the redshift of the source. We modeled the absorption features in our spectra using the model {\it xabs} in SPEX, which calculates the transmission through a slab of material where all the ionic column densities are linked through a photoionization balance model. The latter was provided to {\it xabs} as an input, and it was calculated with CLOUDY\footnote{\tt http://www.nublado.org} v13.01 (\citealt{Fer13}) using as inputs the SEDs shown in Sect~\ref{seds}. For the fits we left as free parameters in {\it xabs} the ionization parameter $\xi$, the hydrogen column density $N_{\rm H}$, the velocity width $\sigma$, and the outflow velocity $v_{\rm out}$. The ionization parameter is a measure of the ionization state of the gas, and it is defined as

\begin{equation}
\label{xidef}
\xi = \frac{L_{\rm ion}}{nR^2},
\end{equation}

\noindent where $L_{\rm ion}$ is the ionizing luminosity in the $1-1000$~Ryd range, $n$ is the hydrogen density of the gas, and $R$ is the distance of the gas to the ionizing source.

As shown in \citet{Ebr16}, the obscuring material that caused the low soft X-ray flux state of the source in 2013 (\citealt{Par14}) was still present in the 2003 observation and, to a lower extent, in the 2015 observations. The obscurer is mildly ionized and its observed variability can be described solely by changes in its covering fraction (\citealt{Ebr16}). We therefore modeled the obscurer with a {\it xabs} component where, in addition to the free parameters mentioned above, we also let the covering fraction $f_{\rm c}$ free.

The fits were carried out with the following approach. We started by adding a {\it xabs} component for the obscurer using the ionzation balance generated with the unobscured SED. Since the obscurer is likely located closer to the central engine than the regular WA, possibly close to the BLR (\citealt{Ebr16}), it is reasonable to assume that it will be irradiated by the unobscured, intrinsic spectrum of the AGN. After this, we added additional {\it xabs} components to model the multi-phase WA on a one-by-one basis until the fit no longer improved. The WA {\it xabs} components were fed with the ionization balance generated with the obscured SED (the SED filtered by the obscurer). The progress of the fit in terms of C-stat and intermediate $\log \xi$ values is shown in Table~\ref{fitprogress}. It can be seen that for the 2015 observations a total of five {\it xabs} components were required, one to account for the obscurer and a WA with four distinct ionization phases. The 2003 and 2013 observations, on the other hand, analyzed in the $7 - 26$~\AA~band due to the obscured continuum at longer wavelengths, only required three and four {\it xabs} components, respectively: one for the obscurer and two or three WA phases. In these two observations the addition of an additional {\it xabs} component only produced a marginal improvement in the fit, $\Delta ({\rm C-stat})/\Delta (\rm d.o.f.) = 7/4$ in 2003 and $\Delta ({\rm C-stat})/\Delta (\rm d.o.f.) = 5/4$ in 2013, respectively. The resulting best-fit parameters are reported in Table~\ref{bestWA}. The RGS spectrum of the second observation in 2015 is shown in Fig.~\ref{spec15b}. The remaining spectra used in this work are shown in Appendix~\ref{rgsspectra}.

\begin{table*}
  \centering
  \caption[]{\label{fitprogress}Progress of the fit after including an additional warm absorber component in each iteration. The values refer to the ionization parameter $\log \xi$ in each {\it xabs} component. The values in bold are those of the obscurer. The underlined values are the adopted best-fit values.}
  \begin{tabular}{c|lc|lc}
    \hline
    \noalign{\smallskip}
    \multicolumn{1}{c}{} & \multicolumn{2}{c}{Obs. 2003 (d.o.f. = 394)} & \multicolumn{2}{c}{Obs. 2013 (d.o.f. = 359)} \\
    \noalign{\smallskip}
    \hline
    \noalign{\smallskip}
    Model & C-stat & $\log{\xi}$ & C-stat & $\log{\xi}$ \\ 
    0{\it xabs} & 840 & - & 645 & - \\ 
    1{\it xabs} & 524 & \textbf{1.15} & 525 & \textbf{0.78} \\ 
    2{\it xabs} & 468 & 2.39, \textbf{1.04} & 474 & 1.96, \textbf{1.61} \\ 
    3{\it xabs} & \underline{\underline{448}} & 2.44, 1.82, \textbf{1.07} & 462  & 1.95, 1.05, \textbf{1.67} \\ 
    4{\it xabs} & $\dots$ & $\dots$  & \underline{\underline{452}} & 2.01, 2.51, 1.01, \textbf{1.68} \\ 
    \noalign{\smallskip}
    \hline
    \noalign{\smallskip}
    \multicolumn{1}{c}{} & \multicolumn{2}{c}{Obs. 2015a (d.o.f. = 854)} & \multicolumn{2}{c}{Obs. 2015b (d.o.f. = 853)} \\
    \noalign{\smallskip}
    \hline
    \noalign{\smallskip}
    Model & C-stat & $\log{\xi}$ & C-stat & $\log{\xi}$ \\ 
    0{\it xabs} & 2939 & - & 3147 & - \\ 
    1{\it xabs} & 1844 & \textbf{1.17} & 1843 & \textbf{1.43} \\ 
    2{\it xabs} & 1610 & 2.63, \textbf{1.18} & 1620 & 2.80, \textbf{1.44} \\ 
    3{\it xabs} & 1135 & 2.82, 2.13, \textbf{0.98} & 1281 & 2.94, 2.13, \textbf{1.34} \\ 
    4{\it xabs} & 1038 & 2.82, 2.56, 1.95, \textbf{0.79} & 1127& 2.91, 2.93, 2.00, \textbf{0.72} \\ 
    5{\it xabs} & \underline{\underline{1028}} & 2.84, 2.58, 2.01, 1.64, \textbf{0.90} & \underline{\underline{1056}} & 2.93, 2.93, 2.04, 1.49, \textbf{0.92} \\ 
    \noalign{\smallskip}
    \hline
    \noalign{\smallskip}
  \end{tabular}
\end{table*}

\begin{table*}
  \centering
  \caption[]{Best-fit values for the obscurer (Component $O$) and WA (Components $A$~to $D$) in NGC 985.}
  \label{bestWA}
  \begin{tabular}{c|c|cccc}
    \hline
    \hline
    \noalign{\smallskip}
    Component & Parameter & Obs. 2003 & Obs. 2013 & Obs. 2015a & Obs. 2015b  \\ 
    \noalign{\smallskip}
    \hline
    \noalign{\smallskip}
    $O$ & $N_{\rm H}$\tablefootmark{a} & $2.6 \pm 0.8 \times 10^{22}$ & $7.8 \pm 1.8 \times 10^{22}$ & $2.8 \pm 0.6 \times 10^{22}$ & $1.3 \pm 0.8 \times 10^{22}$  \\ 
    & $f_{\rm c}$\tablefootmark{b} & $0.62 \pm 0.06$ & $0.94 \pm 0.01$ & $0.49 \pm 0.09$ & $0.18 \pm 0.07$   \\ 
    & $\log \xi$\tablefootmark{c} & $1.07^{+0.04}_{-0.11}$ & $1.68 \pm 0.05$ & $0.90 \pm 0.17$  & $0.92^{+0.42}_{-0.59}$ \\
    & $\sigma$\tablefootmark{d} & $590 \pm 140$ & $1800 \pm 1500$ & $100^{+160}_{-60}$  & $110^{+150}_{-70}$ \\
    & $v_{\rm out}$\tablefootmark{e} & $-6820 \pm 340$ & $-11400 \pm 3200$ & $-6450 \pm 280$ & $-5200^{+600}_{-370}$  \\
    \noalign{\smallskip}
    \hline
    \noalign{\smallskip}
    $A$ & $N_{\rm H}$\tablefootmark{a} & $\dots$ & $1.3 \pm 0.2 \times 10^{22}$ & $1.6^{+3.1}_{-0.8} \times 10^{22}$ & $2.3^{+2.4}_{-1.1} \times 10^{22}$  \\ 
    & $\log \xi$\tablefootmark{c} & $\dots$ & $2.01 \pm 0.05$ & $2.84 \pm 0.03$ & $2.93 \pm 0.03$ \\ 
    & $\sigma$\tablefootmark{d} & $\dots$ & $35^{+30}_{-10}$ & $60^{+55}_{-40}$ & $< 20$  \\ 
    & $v_{\rm out}$\tablefootmark{e} & $\dots$ & $-7200^{+300}_{-60}$ & $-5100 \pm 100$ &  $-5100^{+150}_{-50}$   \\ 
    \noalign{\smallskip}
    \hline
    \noalign{\smallskip}
    $B$ & $N_{\rm H}$\tablefootmark{a} & $2.0^{+1.7}_{-0.7} \times 10^{22}$ & $2.3^{+2.9}_{-1.0} \times 10^{22}$ & $3.6 \pm 1.2 \times 10^{22}$  & $1.0 \pm 0.6 \times 10^{23}$  \\ 
    & $\log \xi$\tablefootmark{c}  & $2.44 \pm 0.08$ & $2.51 \pm 0.06$ & $2.58 \pm 0.06$ & $2.93 \pm 0.02$  \\
    & $\sigma$\tablefootmark{d} & $80^{+60}_{-30}$ & $400 \pm 250$ &  $< 20$ & $30 \pm 10$  \\ 
    & $v_{\rm out}$\tablefootmark{e} & $-210^{+230}_{-170}$ & $-600^{+500}_{-350}$ & $-700 \pm 60$ & $-640^{+100}_{-30}$ \\ 
    \noalign{\smallskip}
    \hline
    \noalign{\smallskip}
    $C$ & $N_{\rm H}$\tablefootmark{a} &  $5.7 \pm 1.5 \times 10^{21}$ & $\dots$ & $4.9 \pm 0.7 \times 10^{21}$ & $5.7 \pm 0.4 \times 10^{21}$  \\ 
    & $\log \xi$\tablefootmark{c} & $1.82 \pm 0.05$ & $\dots$ & $2.01 \pm 0.04$ & $2.04 \pm 0.02$  \\ 
    & $\sigma$\tablefootmark{d} & $60^{+50}_{-30}$ & $\dots$ & $95 \pm 25$ & $125 \pm 15$  \\
    & $v_{\rm out}$\tablefootmark{e} & $-385^{+240}_{-115}$ & $\dots$ & $-365 \pm 45$ & $-350 \pm 30$  \\ 
    \noalign{\smallskip}
    \hline
    \noalign{\smallskip}
   $D$ & $N_{\rm H}$\tablefootmark{a} & $\dots$ & $1.5 \pm 0.5 \times 10^{21}$ & $1.0 \pm 0.6 \times 10^{21}$ & $1.0 \pm 0.3 \times 10^{21}$  \\ 
    & $\log \xi$\tablefootmark{c} & $\dots$ & $1.01 \pm 0.17$ & $1.64 \pm 0.14$ & $1.49 \pm 0.09$  \\ 
    & $\sigma$\tablefootmark{d} & $\dots$ & $60 \pm 40$ & $65^{+80}_{-30}$ & $85^{+45}_{-30}$  \\ 
    & $v_{\rm out}$\tablefootmark{e} & $\dots$ & $-920^{+420}_{-150}$ & $-560 \pm 75$ & $-580 \pm 75$ \\ 
    \noalign{\smallskip}
    \hline
    \noalign{\smallskip}
    \multicolumn{2}{l}{C-stat/d.o.f.}  & $440/394$ & $452/359$ & $1028/854$ & $1056/853$ \\
    \noalign{\smallskip}
    \hline
  \end{tabular}
  \tablefoot{
    \tablefoottext{a}{Hydrogen column density, in units of cm$^{-2}$;}
    \tablefoottext{b}{covering fraction;}
    \tablefoottext{c}{ionization parameter, in units of erg cm s$^{-1}$;}
    \tablefoottext{d}{velocity width, in units of km s$^{-1}$;}
    \tablefoottext{e}{outflow velocity, in units of km s$^{-1}$.}
  }
\end{table*}

\begin{figure*}
  \centering
  \hbox{
  \includegraphics[width=6.5cm,angle=-90]{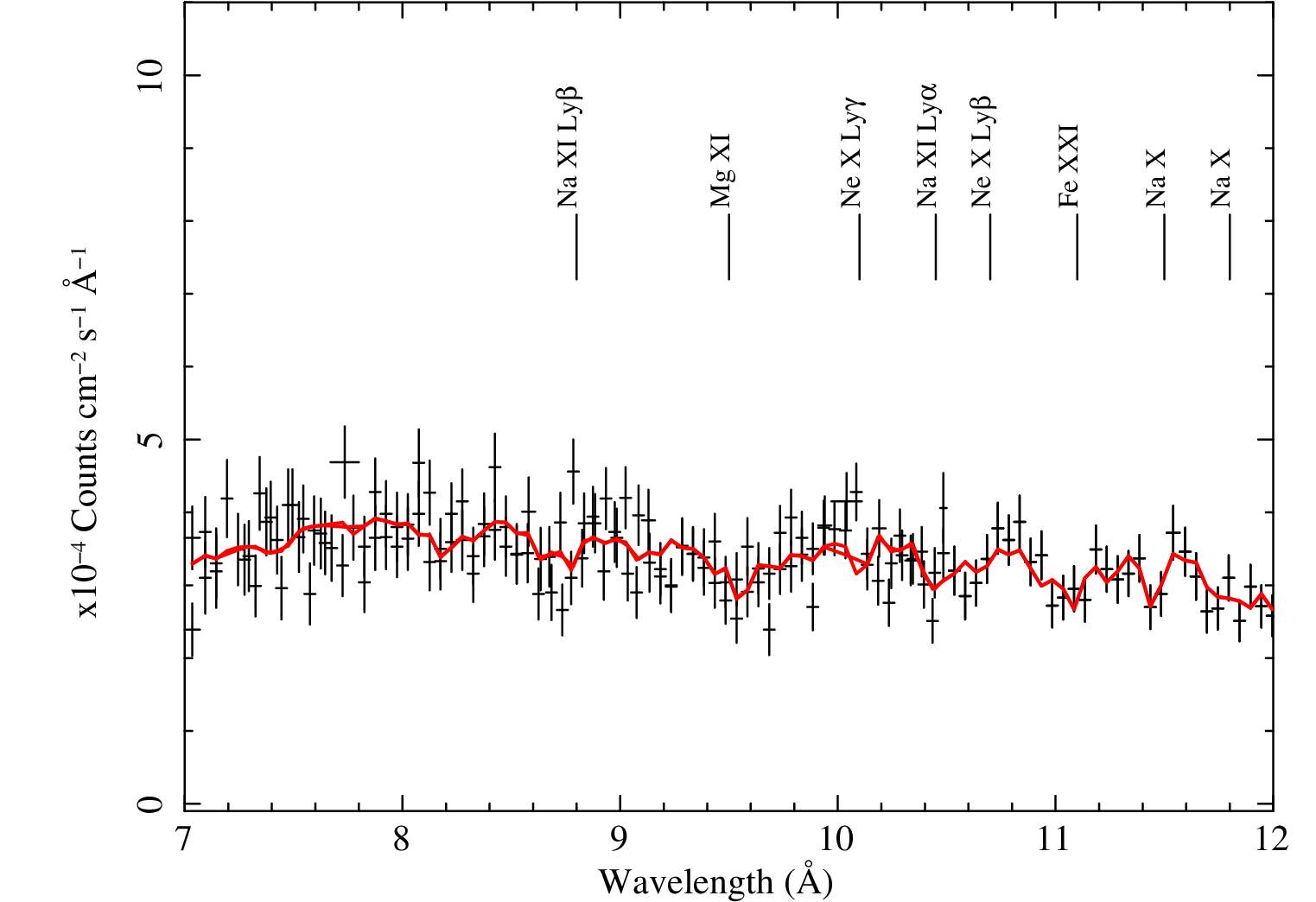}
  \includegraphics[width=6.5cm,angle=-90]{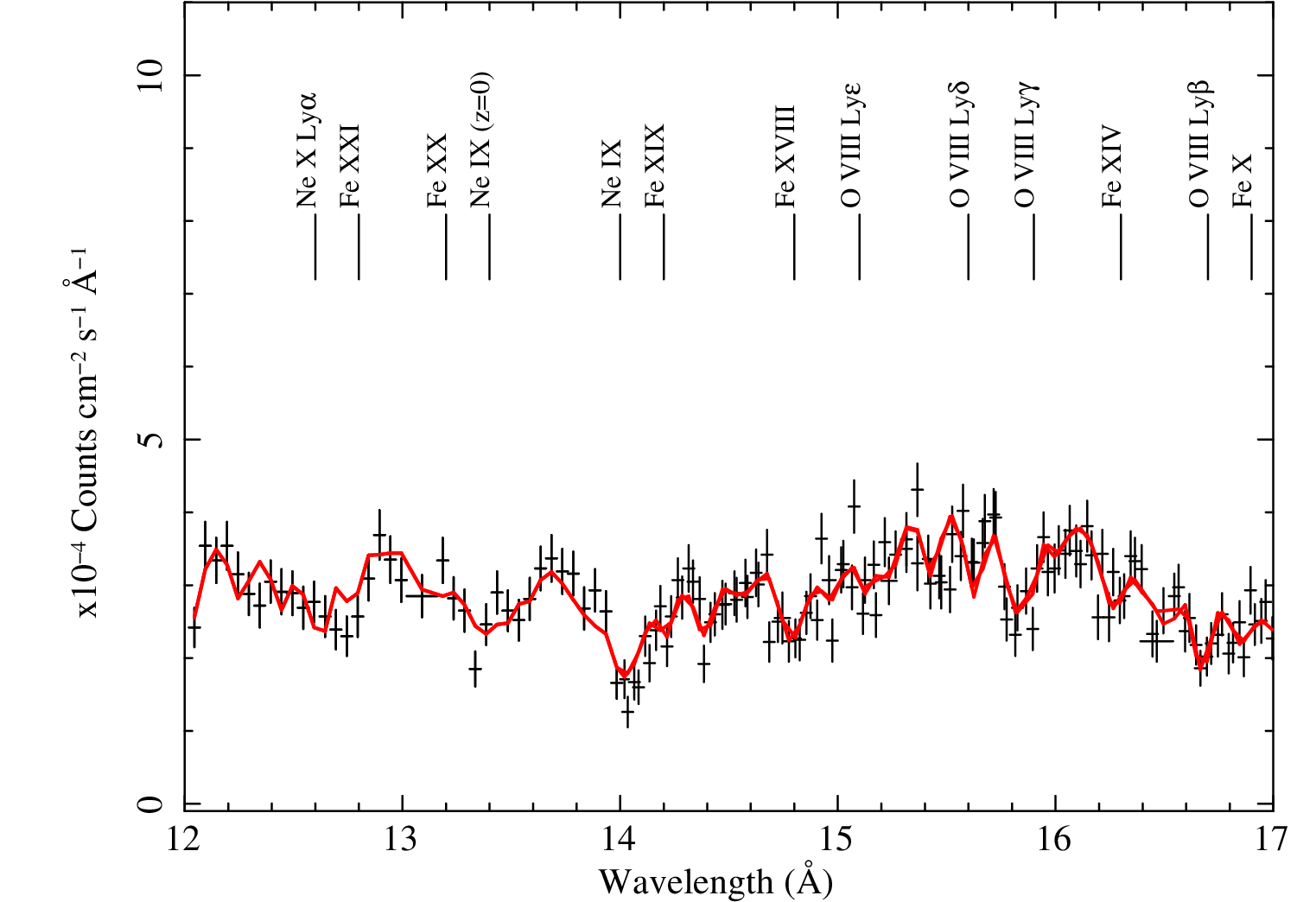}
  }
  \hbox{
  \includegraphics[width=6.5cm,angle=-90]{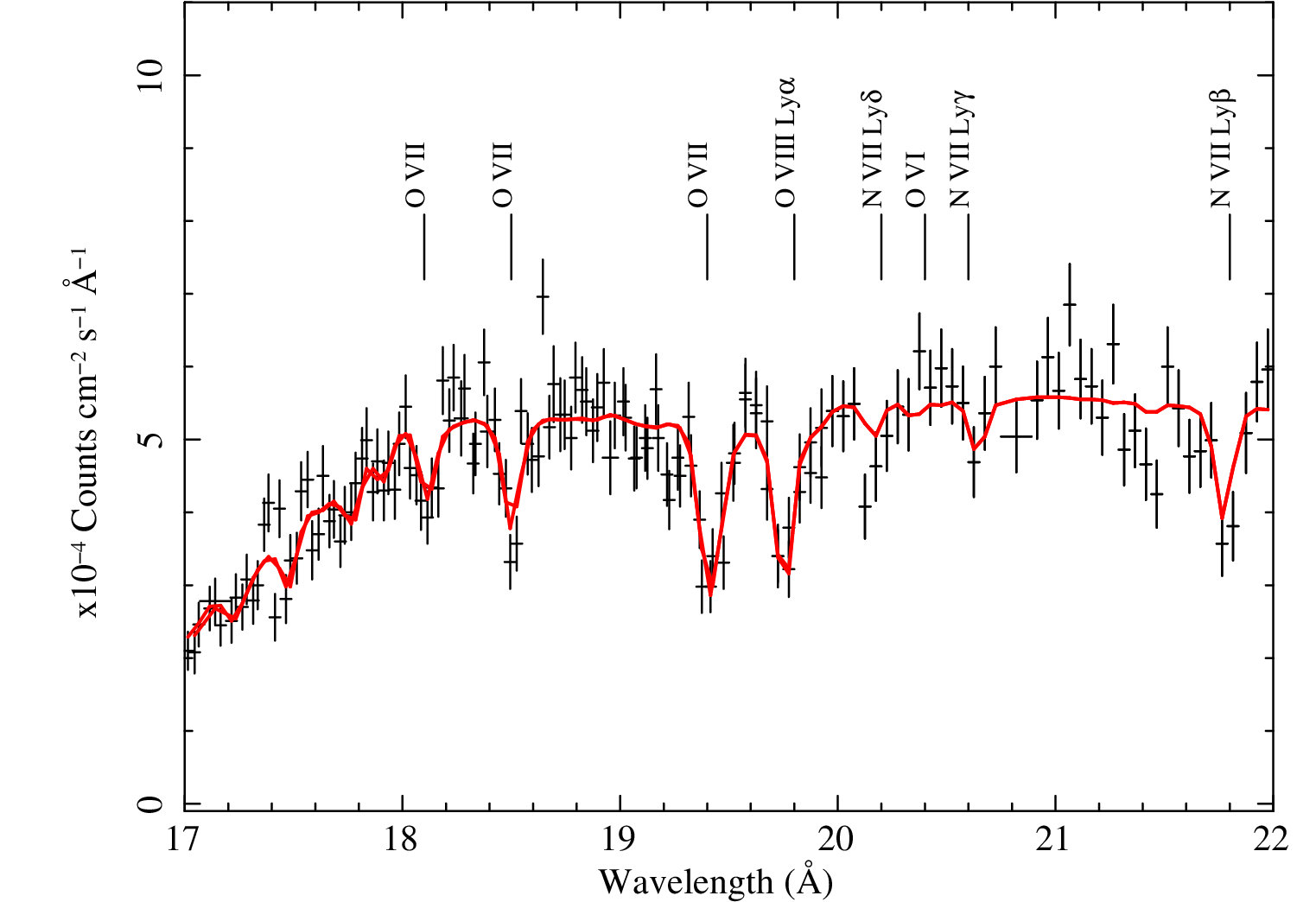}
  \includegraphics[width=6.5cm,angle=-90]{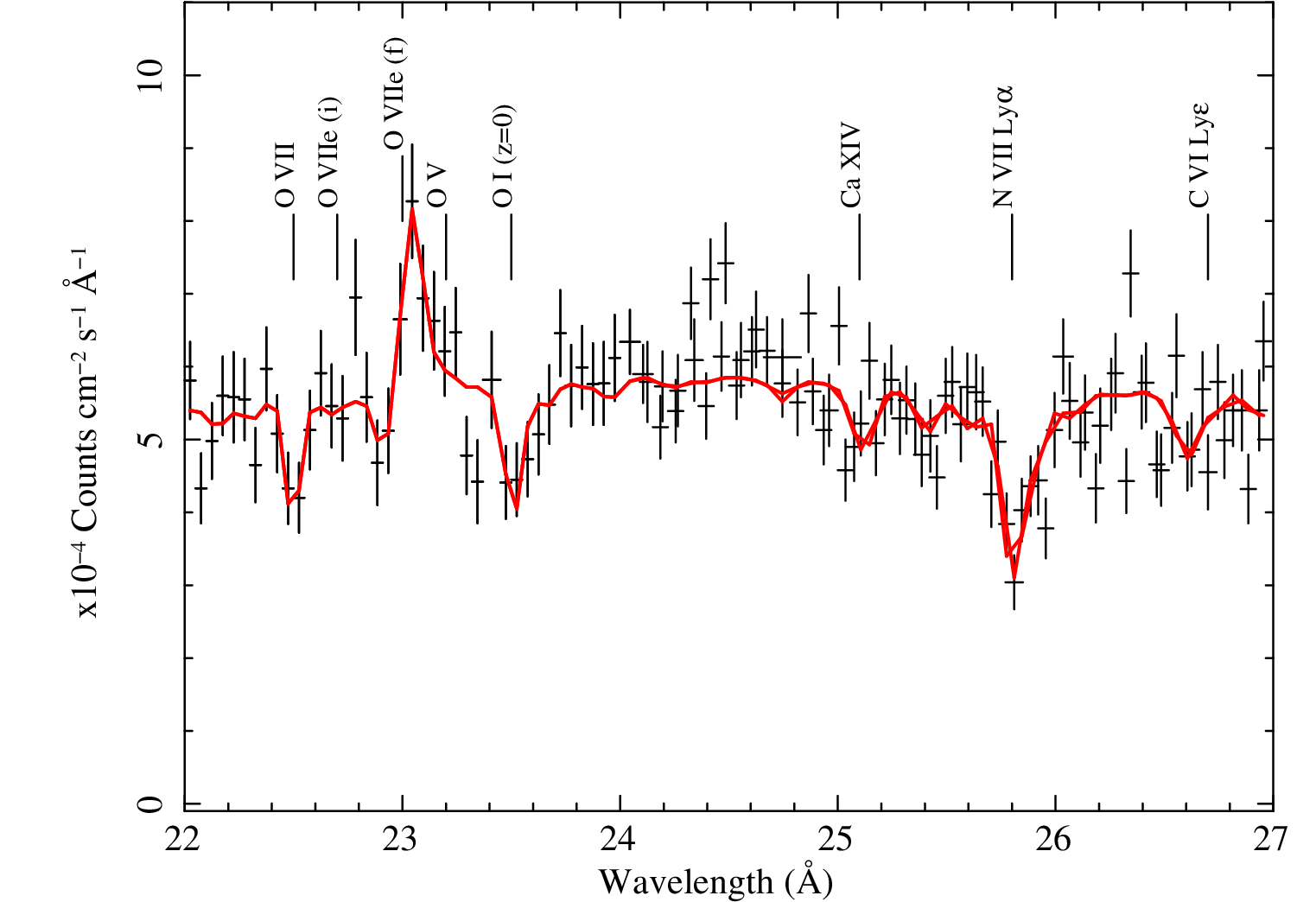}
  }
  \hbox{
  \includegraphics[width=6.5cm,angle=-90]{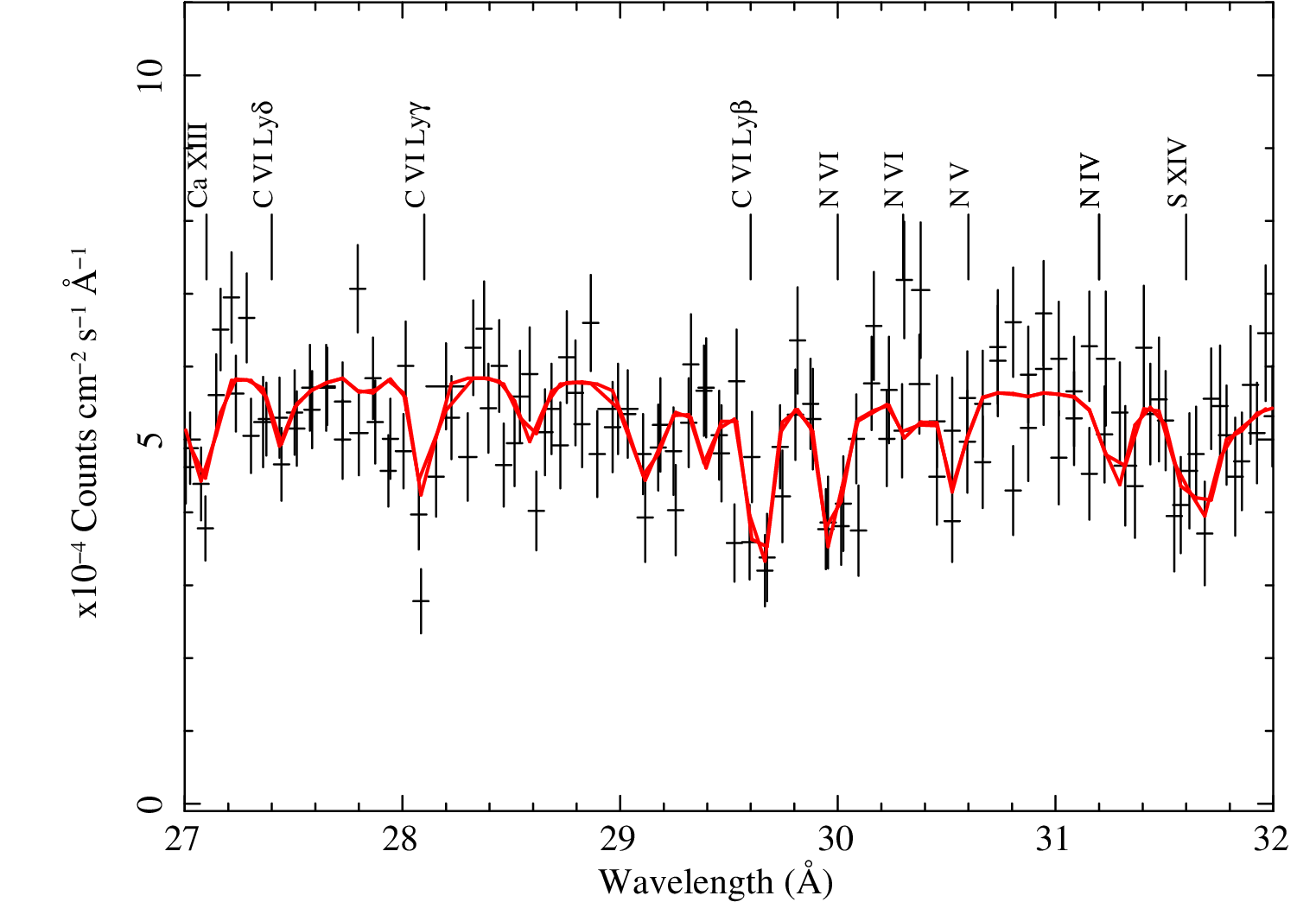}
  \includegraphics[width=6.5cm,angle=-90]{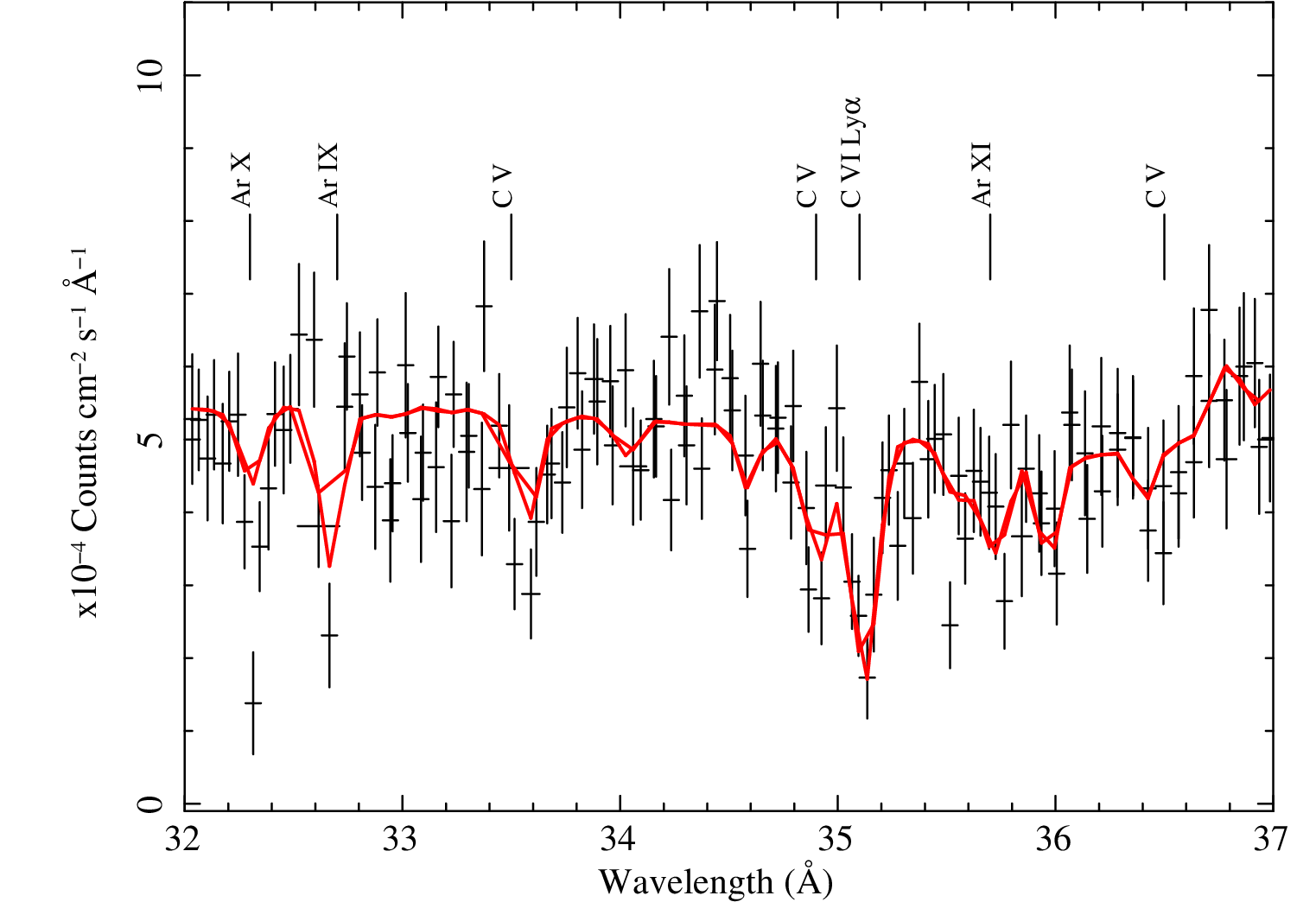}
  }
  \caption{\label{spec15b}\XMM{}~RGS spectrum of NGC 985 in 2015 (second observation). The solid red line represents the best fit model. The most relevant features have been labeled.}
\end{figure*}


\section{Results}
\label{results}

\subsection{The obscurer}
\label{obscurer}

NGC 985 experienced a low soft X-ray flux state in 2013 which was attributed to the passing by of an obscuring cloud of material across our line of sight (\citealt{Par14}). The analysis of the \XMM{}~EPIC-pn spectra of NGC 985 revealed that this obscuration event was possibly recurrent since there were signatures of obscuration in 2003, followed by an unobscured period in 2007-2008 (according to \swift{}~monitoring), before entering in the obscuration event of 2013. Interestingly, the analysis of the 2015 \XMM{}~observations showed that some hints of obscuration were still present in the X-ray and UV spectra of NGC 985 (\citealt{Ebr16}).

\begin{figure}
  \centering
  \includegraphics[width=6.5cm,angle=-90]{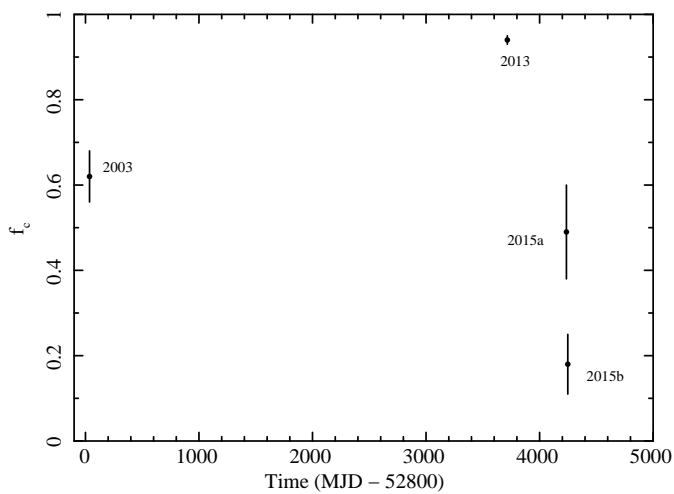}
  \caption{\label{fctimefig}Covering fraction $f_{\rm c}$ of the obscurer in the different \XMM{}~observations.}
\end{figure}

The results of the fits to the RGS spectra shown in Table~\ref{bestWA} reveal that the Hydrogen column density of the obscurer remained constant throughout the observations, with $N_{\rm H} \sim 2 \times 10^{22}$~cm$^{-2}$, with the exception of the 2013 observation, in which the best-fit value is $\sim 3$ times larger. The main consequence is that all the observed changes due to the presence of the obscurer can solely be attributed to variations in the covering fraction of this material. During the monitoring campaign of 2013 the obscurer was almost fully blocking our line of sight ($f_{\rm c}=0.94$), whereas in 2003 the covering fraction was almost two thirds. The inclusion of this obscurer component in the fits of the 2015 observations, when the source had a soft X-ray flux $\sim 10$~times higher than in 2013 and almost twice as bright as in 2003, resulted in a lower covering fraction in the first of the 2015 observations ($f_{\rm c}=0.49$) and a much smaller, but not negligible value in the second observation in 2015 ($f_{\rm c}=0.18$), as shown in Fig.~\ref{fctimefig}.

These covering fraction variations can be interpreted in the context of various possible scenarios. In one of them the obscurer is most likely clumpy so that a fraction of the continuum emission is leaking through it. The different covering fractions observed could indicate that our line of sight is pointing toward the edge of the outflow in the 2003 and 2015 observations, while in 2013 it is intercepting the bulk of the obscuring wind. Another point is related to the duration of the obscuring events. The recurrent appearance and disappearance of the obscurer may indicate that it is transient, although the possibility that it is launched in the form of a continuous stream of gas cannot be ruled out. In the latter case, the intrinsic motion of the wind in the proximity of the central engine would make the outflow intersect our line of sight and drift away from it at the different epochs. An extreme case of this situation would be the long-lasting obscuration event in NGC 5548, which has persistently been in this state for the last several years (\citealt{Meh16}).

\subsection{The warm absorber}
\label{wa}

The soft X-ray spectra of NGC 985 show several absorption troughs associated with intervening ionized winds outflowing at different kinematic regimes. The best-fit results of the higher signal-to-noise \XMM{}~RGS spectra in 2015 revealed that these outflows are composed of four components with distinct ionization states, although the actual values of $\log \xi$ seem to fluctuate between the different observations. This will be discussed in detail in Sect.~\ref{variability}.

All of the WA components show deep Hydrogen column densities $N_{\rm H}$, ranging from $\sim 10^{21}$ to a few times $10^{22}$~cm$^{-2}$. The highest ionization component $A$, with $\log \xi_{\rm A} \sim 2.9$, is also the fastest, outflowing at $\sim -5\,100$~km s$^{-1}$. The slowest WA component $C$, with $\log \xi_{\rm C} \sim 2.0$, is moving at $\sim -350$~km s$^{-1}$, while the remaining components $B$ ($\log \xi_{\rm B} \sim 2.6$) and $D$ ($\log \xi_{\rm D} \sim 1.6$) are kinematically more difficult to disentangle, both traveling at around $-650$~km s$^{-1}$ and $-560$~km s$^{-1}$, respectively, with overlapping error bars. The measured turbulent velocity $\sigma$ is also mostly consistent across observations. In some cases, such as components $A$ and $C$ in 2013, $\sigma$ appears to adopt higher values but with very large associated uncertainties.

The observations of 2003 and 2013, affected by the shorter exposure times and the flux supression due to the presence of the obscurer, had a worse signal-to-noise ratio and could only be analyzed in the $7-26$~\AA~band. Only two WA components could be significantly detected in 2003, and three in 2013. Their cross-identification with the WA phases in the 2015 observations was not straightforward in some cases. For instance, for the two WA components detected in the 2003 observation, the one with $\log \xi = 1.82$ outflowing at $-385$~km s$^{-1}$ was immediately cross-identified with component $C$ in the 2015 observations since its kinematic properties and column density matched well those of 2015, albeit with a somewhat lower ionization parameter. The other WA component was identified with component $B$ based on the similar column density and ionization parameter as those in 2015, although their kinematics were difficult to reconcile with those of that component in 2015. The measured outflow velocity of $-210$~km s$^{-1}$ was smaller than the outflow velocity measured in 2015 for this component, $\sim -650$~km s$^{-1}$, but the large error bars in the former make these values consistent within less than $2\sigma$.

The three components detected in the 2013 observation, the most affected by the obscuration, were also difficult to cross-identify with their counterparts in 2015 in some cases. Two of them were identified with component $A$ and $B$, based on their similarities in kinematics and column densities. The third component was identified as component $D$, based on the measured column density. It is worth noting that all the detected components in the 2013 observations seem to be kinematically faster with respect to their counterparts in 2015. The measured outflow velocity in components $B$, $D$, and in the obscurer itself have such large uncertainties that are consistent within $1\sigma$ with the velocities seen in 2015. The only exception is component $A$, which seems to be significantly faster in 2013. Typical accelerations (or decelerations) due to radiation pressure at those distances make it difficult to reconcile such values. While the possibility that we are detecting another, different kinematic component is also a reasonable explanation for this disparity, the fact that all components seem to be faster (albeit the large uncertainties in most of the cases) may indicate that some other systematic process is in action. A change in the projected outflow velocity may also indicate a change in the angle subtended between the outflow and our line of sight. However, for a change of about $30\%$ in velocity, if solely explained by this, the angle of the outflow may have changed by almost 45 degrees with respect to the line of sight. Such a bent is somewhat extreme, and it seems difficult to explain how can it happen in timescales of tens of months. The poor statistics, however, do not allow us to explore any of these scenarios in further detail. Furthermore, in all cases, the measured ionization parameter in 2013 is significantly lower than those of their 2015 counterparts, possibly indicating a change in the ionization properties of the WA in response to the dramatic suppression of the ionizing flux (see Sect.~\ref{variability}).


\section{Discussion}
\label{discussion}

\subsection{Thermal stability}
\label{stability}

\begin{figure*}
  \centering
  \hbox{
  \includegraphics[width=6.5cm,angle=-90]{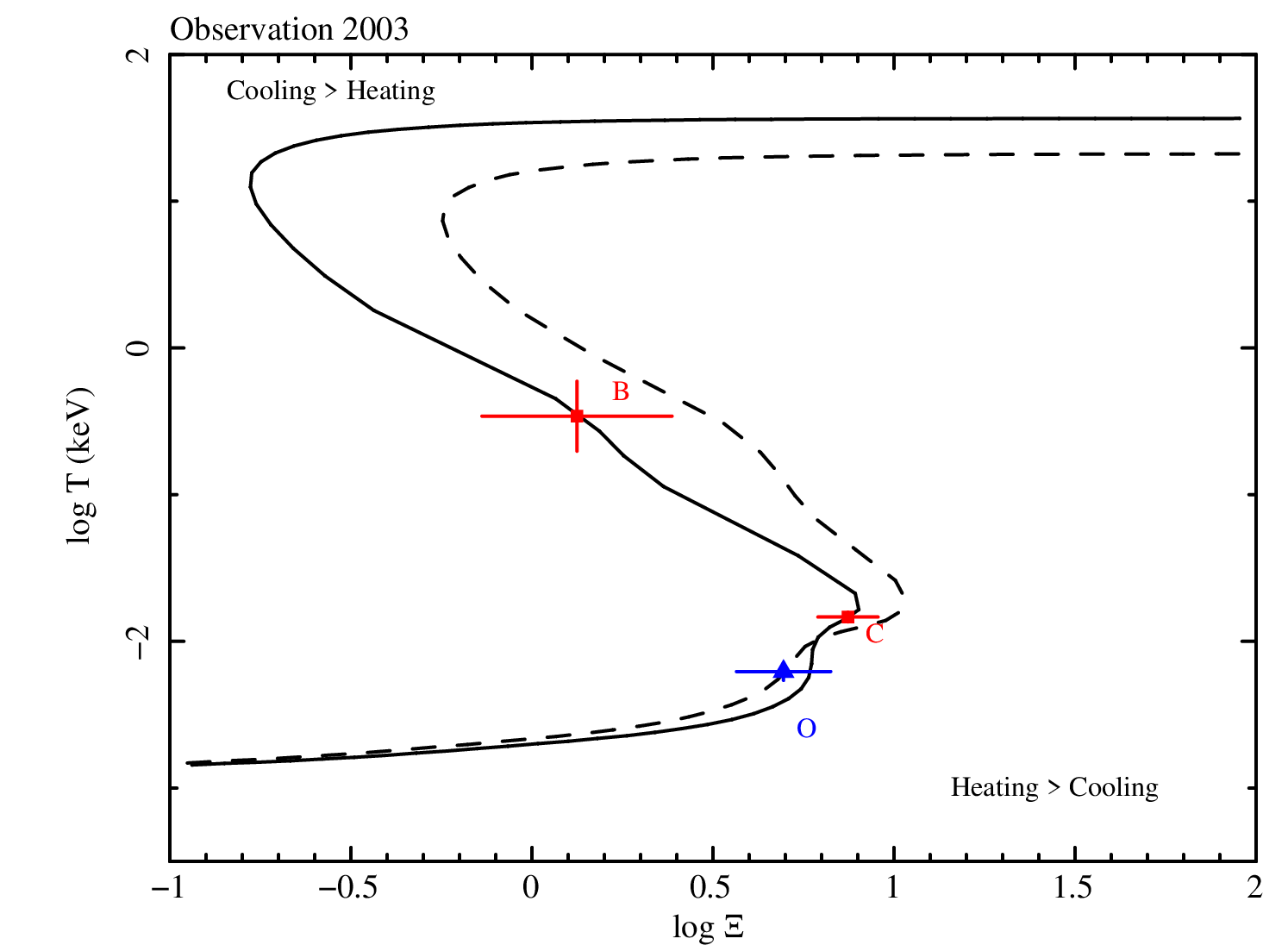}
  \includegraphics[width=6.5cm,angle=-90]{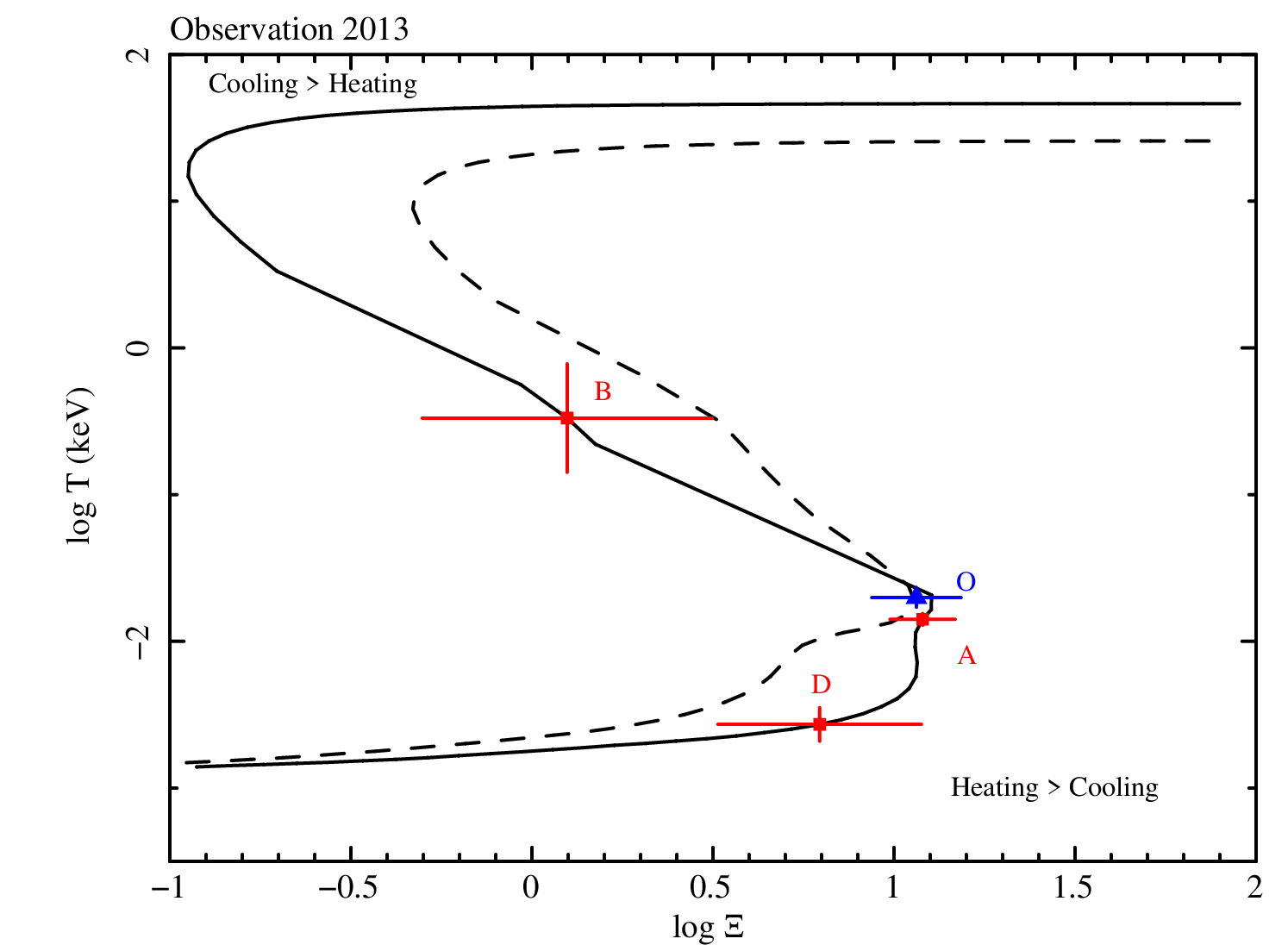}
  }
  \hbox{
  \includegraphics[width=6.5cm,angle=-90]{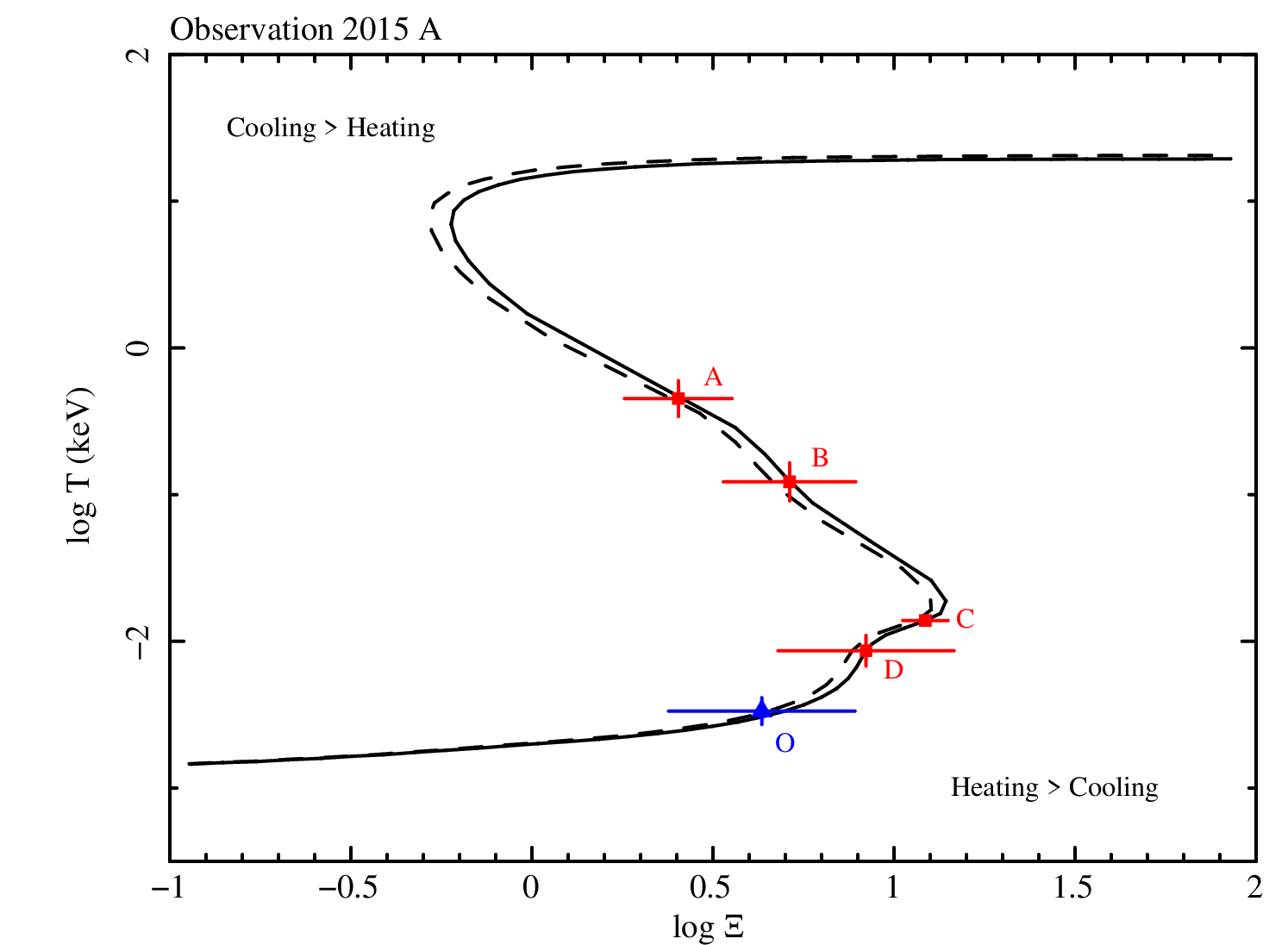}
  \includegraphics[width=6.5cm,angle=-90]{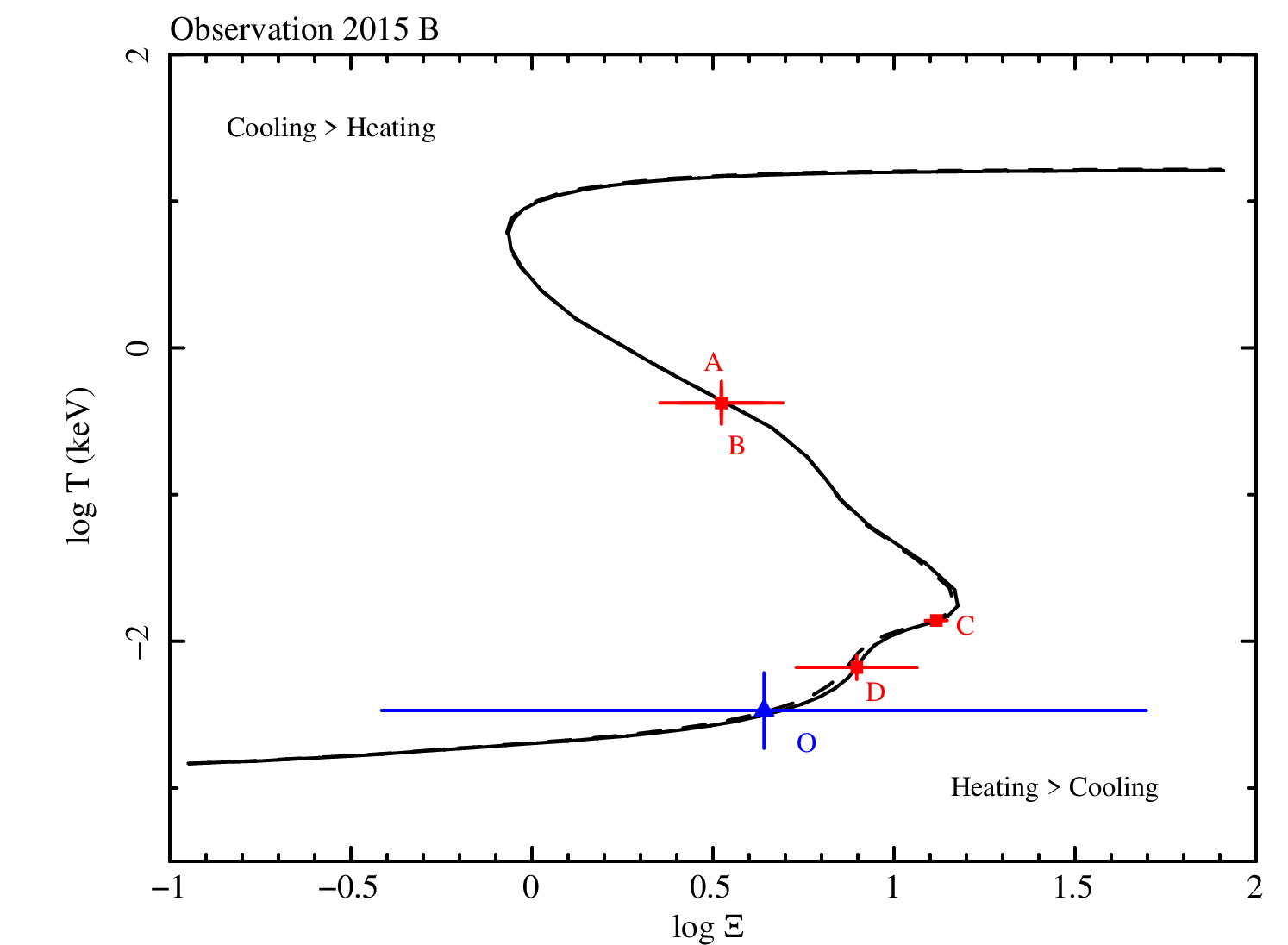}
  }
  \caption{\label{scurves}Pressure ionization parameter as a function of the electron temperature for the unobscured (dashed line) and obscured (solid line) SEDs in the 2003 ({\it top left panel}), 2013 ({\it top right panel}), and the 2015 observations ({\it bottom panels}). The X-ray WA components are represented with filled red squares ($A$~to~$D$) and the obscurer ($O$) is represented with a filled blue triangle.}
\end{figure*}

The thermal stability curves of a photoionized gas, also known as cooling curves or S-curves, provide some insight into the structure of the absorbers. The shape of these curves is determined by the SED that illuminates the gas, and they usually represent the pressure ionization parameter $\Xi$ as a function of the electron temperature $T$ (\citealt{Kro81}). The pressure ionization parameter is defined as:

\begin{equation}
\label{Xidef}
\Xi = L/4\pi r^2cp = \xi/4\pi ckT,
\end{equation}

\noindent where $\xi$ is the ionization parameter as defined in Eq.~\ref{xidef}, $c$~is the speed of light, $k$~is the constant of Boltzmann, and $T$~is the electron temperature. We computed the different values of $\Xi$ in each of the four \XMM{}~observations by feeding CLOUDY v13.01 with their corresponding SEDs (described in Sect.~\ref{seds}) and assuming the proto-Solar abundances of \citet{LP09}. The output was a grid of ionization parameters $\xi$ and their corresponding temperature $T$ for a thin layer of gas irradiated by the ionizing continuum. These curves, shown in Fig.~\ref{scurves}, divide the $\Xi - T$ plane in two regions: above the curve cooling dominates heating, while below the curve heating dominates cooling. Over the points on the curve the cooling rate equals the heating rate, and the gas is therefore in thermal equilibrium. The branch of the curves with negative derivative $d\Xi/dT < 0$, where the curve turns backwards, represent a region in the $\Xi - T$ space which is unstable against isobaric perturbations. As a result, a parcel of gas located in this area that is subject to a small positive temperature perturbation will see its temperature to increase until it reaches the next stable branch (those with positive derivative $d\Xi/dT > 0$). In turn, a negative temperature perturbation will lead to a temperature decrease until the parcel of gas falls to a stable branch.

We have overplotted the WA components, as well as the obscurer, detected in each of the observations over their respective stability curve. The WA components are shown in Fig.~\ref{scurves} as red squares, and the obscurer is represented by a blue triangle. Owing to its mild ionization level, the obscurer is located in the stable branch at the bottom of the curve in all of the observations, sharing this space with component $D$ and, to a lesser extent, with Component $C$. This component $C$, when detected, is located at the edge of this branch, just before the curve turns backwards. Component $B$, on the other hand, is always located in the unstable branch of the curve, changing its position slightly as its ionization state changes in the different epochs. Component $A$ is co-located with component $B$ in the 2015 observations. In 2013, when the source was most obscured, its ionization parameter decreased by almost 1 dex, and thus this component slid down the curve to the stable branch. Unfortunately, this component was not detected in the 2003 observation and therefore we cannot know its behavior prior to the obscured phase.

Stability curves are also useful to probe whether the different ionized phases are in pressure equilibrium, when they share the same $\Xi$. For these observations, however, because of the large uncertainties in several parameters, this kind of analysis is somewhat difficult. Nevertheless, some rough conclusions might be drawn from these plots. The WA components seem to be scattered over the plots but may be divided in two blocks, one with the highest ionization components $A$ and $B$, and another one with the less ionized components $C$ and $D$. Each of these blocks are in principle not in pressure equilibrium with each other, while within each block the pairs of components are within their respective errors. For example, components $A$ and $B$ are both quite close to each other on the unstable branch of the S-curve in the 2015 observations. The exception to this was in the 2013 observation when component $A$ was signficantly less ionized. On the other hand, components $C$ and $D$ share a similar locus in the stable branch, together with the obscurer.

This broad division is at odds with the results of \citet{Kro05}~and \citet{Kro09}. They found that the high- and low-ionization phases (that can be assimilated to each of the blocks described here) were in pressure equilibrium. Their main conclusion is that this components were co-located, or embedded one into another, or that they both were embedded in a third, nondetected, component. This discrepancy can be explained by the different shapes of the stability curves between those works and this one. \citet{Kro05} used \beppo{}~and \chan{}~data from 1999 and 2002, when NGC 985 was only mildly obscured (see Fig. 6 in \citealt{Ebr16}), and \citet{Kro09} used the \XMM{}~2003 observation. If the obscurer is not taken into account when creating the SED, together with the different performances of photoionization codes over time (\citealt{Ebr16b}), this may lead to a different S-curve shape and a different location of the components on it which can explain the discrepancy.

\begin{figure*}
  \centering
  \hbox{
  \includegraphics[width=6.5cm,angle=-90]{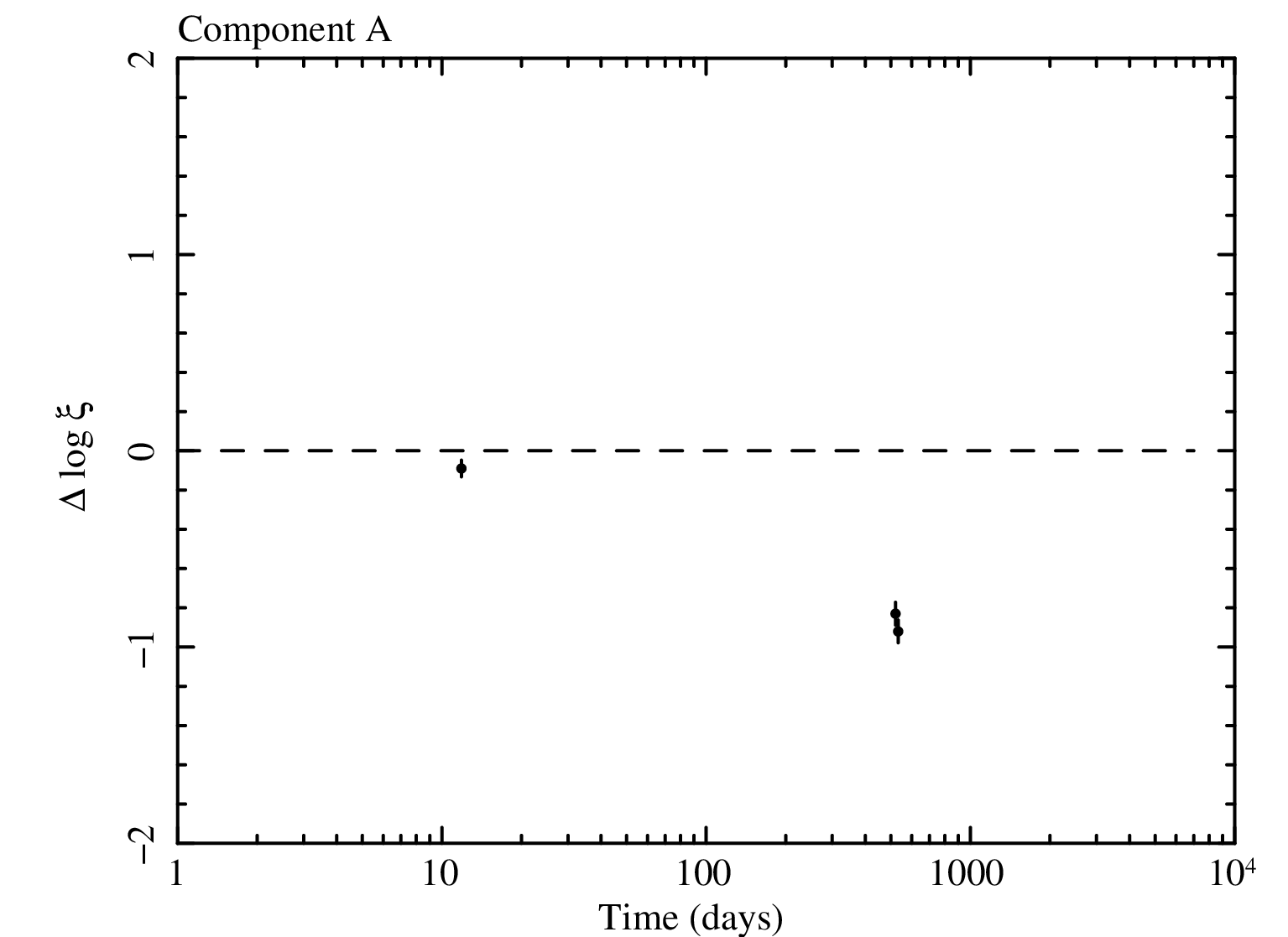}
  \includegraphics[width=6.5cm,angle=-90]{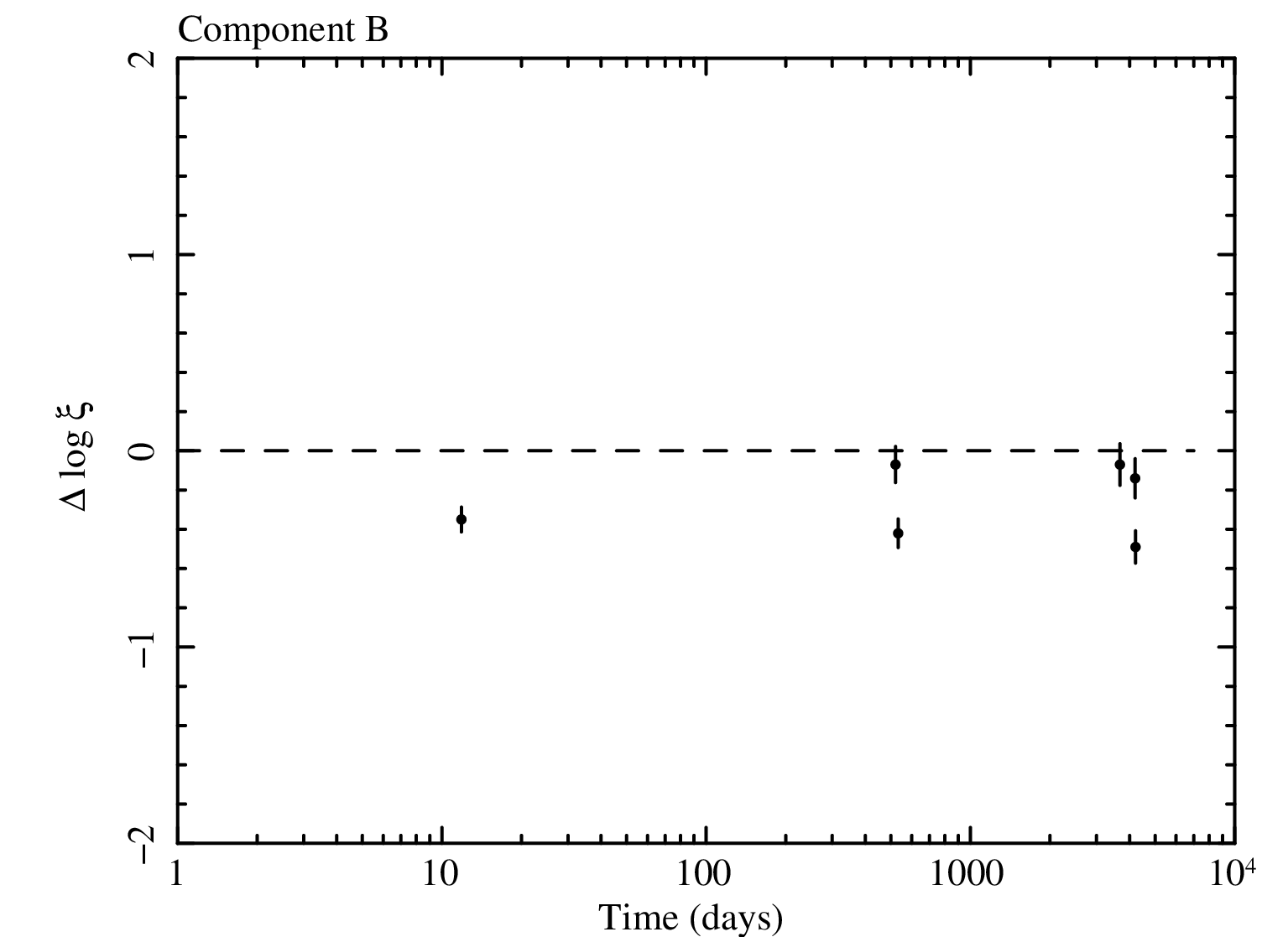}
  }
  \hbox{
  \includegraphics[width=6.5cm,angle=-90]{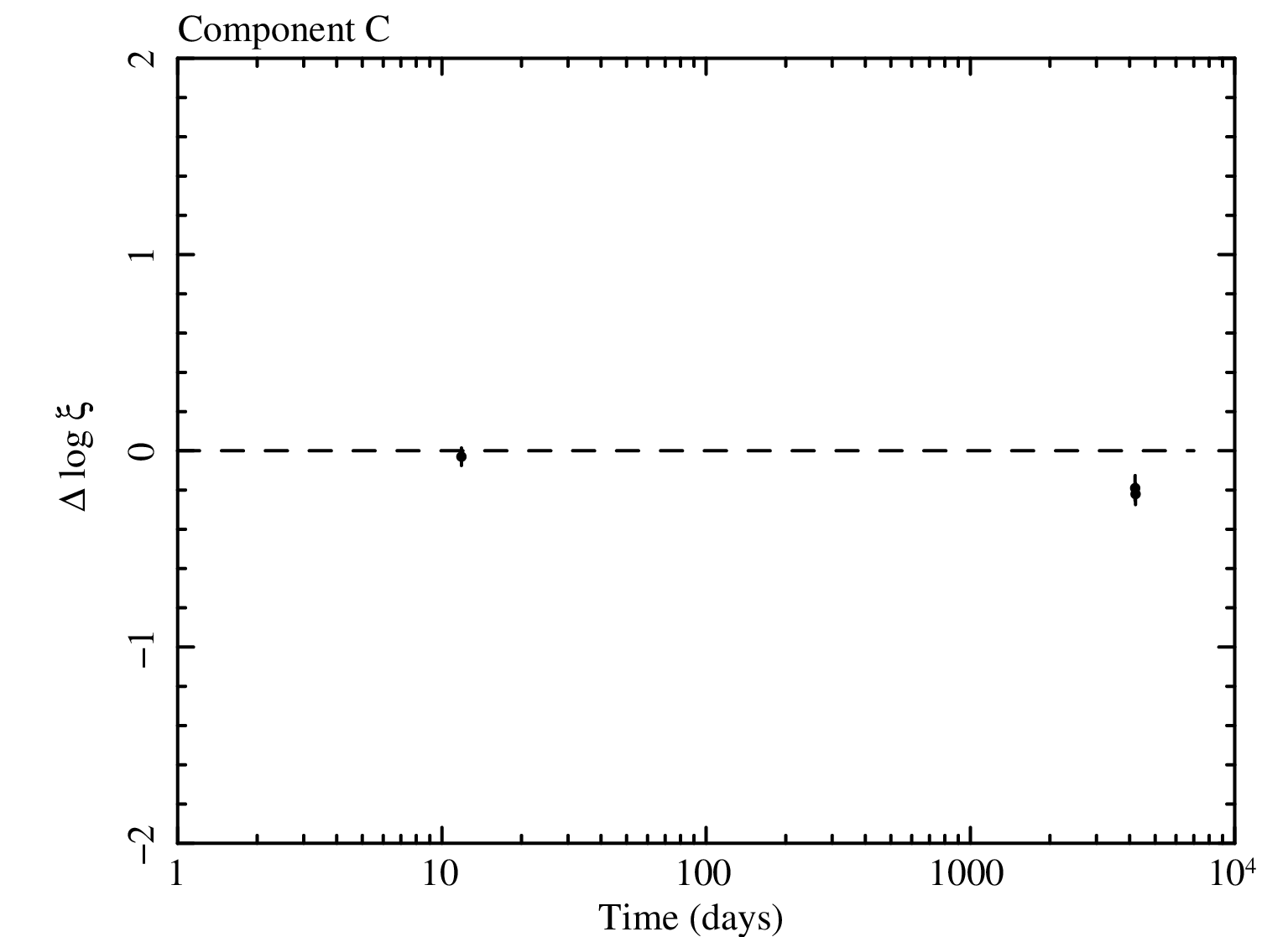}
  \includegraphics[width=6.5cm,angle=-90]{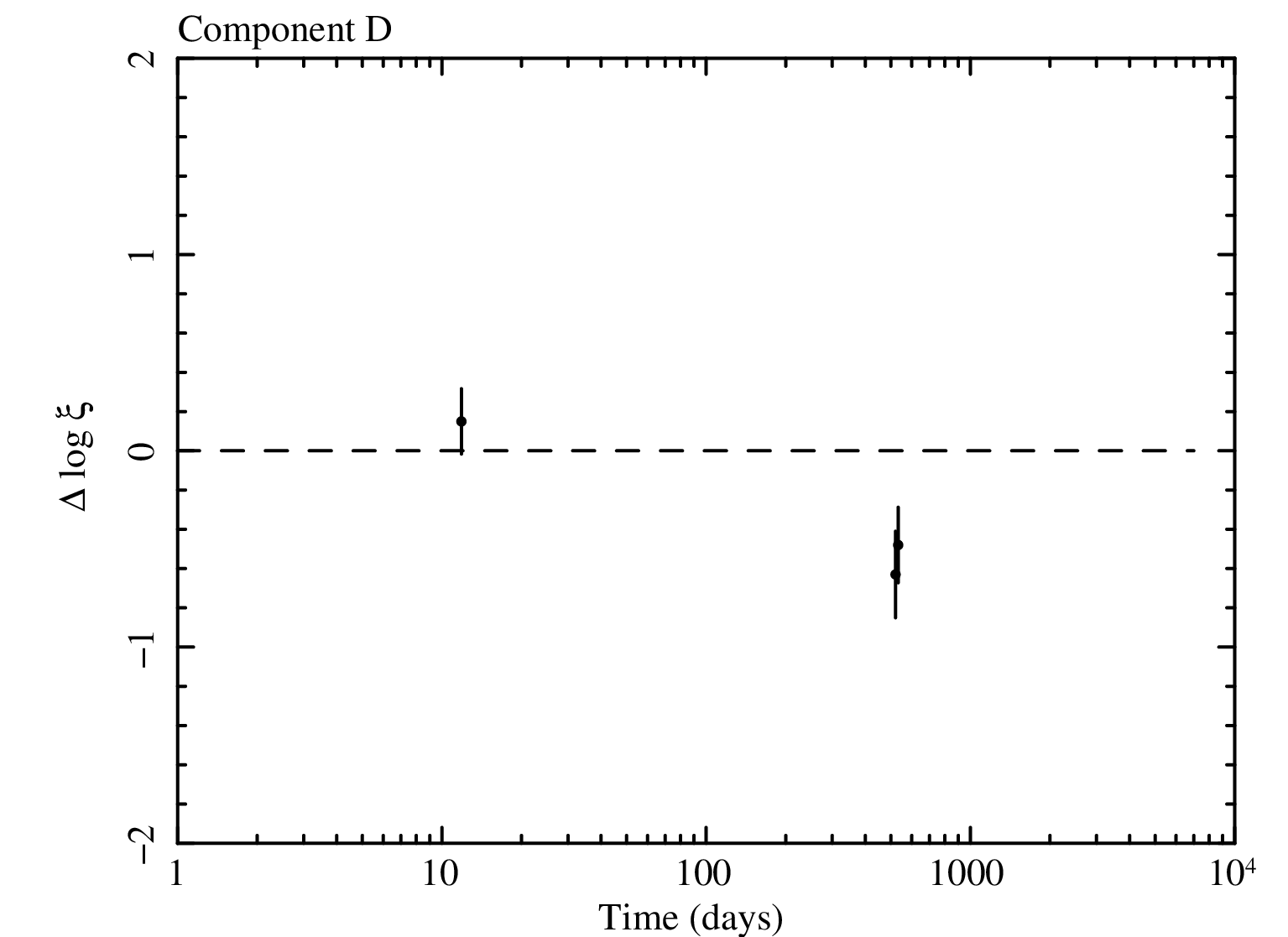}
  }
  \caption{\label{varplots}Variation in the ionization parameter $\Delta \log \xi$ at different time scales (in days) for WA components $A$ and $B$ (top panels), and $C$ and $D$ (bottom panels).}
\end{figure*}

Interestingly, from the results in Table~\ref{bestWA} and the plots in Fig.~\ref{scurves} it is immediately noticeable the similarity in the kinematics between the obscurer and the WA component $A$, both outflowing at comparable velocities, and in their column densities. They also share similar $\Xi$ values, consistent within $1\sigma$, in the S-curves which would mean they are pressure equilibrium, being almost co-located in 2013, when the source was most obscured. The question that arises is whether both absorbers are part of the same long-lived structure. If they are, a plausible scenario that can explain the changes in component $A$ and in the obscurer is that the latter is confined by the former. In this case, when the flux drops component $A$ becomes less ionized and the pressure dramatically decreases, as seen in the 2013 observation. Since this component is in an unstable branch of the curve a perturbation that lowers its temperature would make it travel down the curve until a stable branch is reached. To compensate for this pressure drop, the obscurer must expand, lowering its density and, in turn, increasing its ionization parameter (\citealt{Kro05}). As the flux rises back again during the 2015 observations the process reverts. While this explains the observed behavior, it requires the components to be in ionization equilibrium; otherwise the components would not even need to be located on the stability curve. Furthermore, the different velocity widths may indicate that the obscurer and component $A$ are not co-located (but also see Sect.~\ref{location}). From the UV analysis, given that only $\sim 25$\% of the UV emitting area is obscured while the X-ray emission almost fully obscured, it is likely that the obscurer is much closer to the central source, possibly in or near the BLR (\citealt{Ebr16}; Kriss et al., in preparation).

\subsection{Variability}
\label{variability}

If the changes observed in the WA components are due to photoionization and recombination of the ionized gas in response to the ionizing flux variations, a lower limit on the density of the absorbing gas can be estimated using (\citealt{Bot00}):

\begin{equation}
\label{trec}
t_{\rm rec}(X_i)=\left(\alpha_r(X_i)n\left[\frac{f(X_{i+1})}{f(X_i)}-\frac{\alpha_r(X_{i-1})}{\alpha_r(X_i)}\right]\right)^{-1},
\end{equation}

\noindent where $t_{\rm rec}(X_i)$~is the recombination time scale of a given ion $X_i$, $\alpha_r(X_i)$ is the recombination rate from ion $X_{i-1}$ to ion $X_i$, and $f(X_i)$ is the fraction of element $X$ at the ionization level $i$. The recombination rates $\alpha_r$ of the different ions are known from the atomic physics, while the fractions $f$ can be determined from the ionization balance of the source. Therefore, for a given ion $X_i$ it is possible to obtain a lower limit on the density $n$ if an upper limit on the recombination time $t_{\rm rec}$ is known, since $t_{\rm rec} \propto 1/n$. One has to keep in mind, however, that the estimations obtained with the formula above will hold under the assumption that the gas has reached equilibrium again.

The observations analyzed in this paper allow us to probe time scales that range between 12 days to almost 12 years. We followed a similar approach as in \citet{Ebr16b}, and we calculated the measured variation in the ionization parameter $\xi$ between each individual observation. In this way we can explore the typical time scales at which the different WA components begin to vary. The sampling of the observations is, however, sparse. We cover an ample range of time scales, but at the expense of lacking enough high-resolution spectra at intermediate epochs. This is of course unavoidable, as the investment in observing time for one AGN would be prohibitive otherwise. Each of these observations is a snapshot in the behavior of the ionized absorbers. The observed changes are then assumed to have happened some time between observations, but when exactly is uncertain. The results of this exercise are shown in Fig.~\ref{varplots}. It can be seen that component $A$ begin to significantly vary at time scales of $\sim 520$~days. Component $B$, on the other hand, already showed signs of variability in the $\sim 12$~days spanned between the 2015 observations. Components $C$ and $D$ show some variability, but below the $2\sigma$ level, at around 4200 and 520 days.

The fraction of a given element $X$ in an ionization state $i$, $f(X_i)$ in Eq.~\ref{trec}, can be determined from the ionization balance of NGC 985. It provides the ionic column densities of the different elements that can be divided by the toal hydrogen column density to obtain $f(X_i)$. We identified the ions that contributed most to each WA component based on their column density, oscillator strength, and significance of their spectral features, so that they are the drivers of the fit. We then passed the ionic column densities of these ions to the SPEX auxiliary program {\it rec\_time}, that provides as output the product $nt_{rec}$ for each of them. The tables with the ions used as a proxy for each WA component and their $nt_{rec}$ product are listed in Appendix~\ref{varions}.

With the average $<nt_{rec}>$ for each component, and using the estimated time scales for variability (upper limits on $t_{rec}$) and nonvariability (lower limits on $t_{rec}$) described above, we can obtain lower and upper limits on the density of the absorbing gas, respectively, for each component. As expected, the component that changes the fastest is the one with the highest density, component $B$, with a density higher than $\log n = 4.8$~cm$^{-3}$. The remaining components can have densities as high as $\log n \sim 4.9$~cm$^{-3}$, based on their nonvariability. Among them, component $C$ is possibly the less dense gas since it only shows variability on time scales of $\sim 12$~years. These values are shown in Table~\ref{dens}.

\subsection{Location of the absorbers}
\label{location}

\begin{table*}
  \centering
  \caption[]{Variability time scales, density, and location of the WA components in NGC 985.}
  \label{dens}
  \begin{tabular}{l c c c c c c c}
    \hline\hline
    \noalign{\smallskip}
    Component      & $<nt_{\rm rec}>$  & $t_{\rm rec}^{\rm lower}$  & $t_{\rm rec}^{\rm upper}$  & $\log n^{\rm lower}$  & $\log n^{\rm upper}$  & $R^{\rm lower}$  & $R^{\rm upper}$  \\
           & (s\,cm$^{-3}$) & (days) & (days) & (cm$^{-3}$) & (cm$^{-3}$) & (pc) & (pc) \\
    \noalign{\smallskip}
    \hline
    \noalign{\smallskip}
    $A$ & $7.0 \times 10^{10}$  & $> 12$ & $< 520$  & $> 3.2$  & $< 4.9$ & $> 1$  & $< 7$  \\
    $B$ & $6.9 \times 10^{10}$  & $> 0$ & $< 12$  & $> 4.8$  & $\dots$ & $\dots$  & $< 2$  \\
    $C$ & $5.1 \times 10^{10}$  & $> 12$   & $< 4200$   & $> 2.2$ & $< 4.7$ & $> 3$ & $< 60$  \\
    $D$ & $4.9 \times 10^{10}$  & $> 12$   & $< 520$   & $> 3.0$ & $< 4.7$ & $> 5$ & $< 30$  \\
    \noalign{\smallskip}
    \hline
  \end{tabular}
\end{table*}

From the definition of the ionization parameter in Eq.~\ref{xidef}, the distance of the ionized gas to the central source is given by:

\begin{equation}
\label{disteq}
R = \sqrt{\frac{L_{ion}}{n\xi}}. 
\end{equation}

\noindent The ionizing luminosity of NGC 985 was calculated by integrating the obscured SED between 1 and 1000 Ryd for all observations and then averaging to obtain a value of $4.7 \times 10^{44}$~erg s$^{-1}$. This is the ionizing flux that is seen by the WA components after being filtered by the obscurer. From Eq.~\ref{disteq} and using the upper and lower limits on the gas density derived in Sect.~\ref{variability}, we can obtain lower and upper limits on the distance at which the WA components lie. The results are summarized in Table~\ref{dens}.

The denser gas, faster to respond to flux changes, in component $B$ is located at $\sim 2$~pc or less from the ionizing source. Since we did not probe variability (or lack of) at time scales smaller than 12 days, we cannot set a lower limit on its location. Component $A$ is located between 1 to $\sim 7$~pc from the central SMBH. It is therefore unlikely that this WA component and the obscurer share the same {\it locus}, as discussed in Sect.~\ref{stability}. Component $C$ responds at much longer time scales which would mean that the gas is less dense. For this component we derive the farthest upper limit on the distance, $\lesssim 60$~pc, although it can also be located as close as 3~pc from the source. Finally, component $D$ is likely located between 5 and 30 pc.

To put these values into context we can use the relation in \citet{Kas05} to estimate the location of the BLR in NGC 985. We obtain $R_{BLR} \sim 16$~light-days, or $\sim 0.013$~pc, so it is clear that all the WA components are outflowing much further away. The inner side of the putative dusty torus around the central engine can be approximated by the dust sublimation radius, which is $\sim L_{ion, 44}^{1/2}$~pc (\citealt{KK01}). For our measured $L_{ion}$, this value is $\sim 2.5$~pc. The picture that arises from these values is somewhat similar to what was seen in the stability curves (Fig.~\ref{scurves}). Components $A$~and $B$ are possibly located at pc-scale distances, with $B$ likely closer to the central SMBH but still far from the BLR. They have similar ionization states, but their kinematics are substantially different. Having ruled out that the much faster component $A$ is related to the obscurer, it is possible that both components are powered by different acceleration mechanisms. On the other hand, components $C$~and $D$ may extend up to several tens of pc from the ionizing source. Their location and density are consistent with the typical values of the gas in the NLR of AGN (see e.g., \citealt{Whe15}), although their lower limits may bring them closer to the location of components $A$ and $B$. 

It is worth noting that, in principle, a similar approach to estimate the variability scale, the limits on the density, and ultimately the location could be followed also for the obscurer. However, the obscurer has additional constraints as it is clearly observed in the HST UV data. Given the line widths and covering fractions observed in the UV, the obscurer must be located within the BLR (\citealt{Ebr16}; Kriss et al., in prep.). Constraining the location of the obscurer in X-rays by means of variability changes that provide results compatible with the UV observations would require a much shorter sampling (typically much less than a day). The available datasets, particularly the highly obscured ones in which the obscurer is almost fully covering our line of sight, lack the required signal-to-noise and exposure times to perform this kind of analysis reliably. Therefore in what follows, we rely on the constraints from the UV observations and assume that the obscurer is located within 16 light-days from the central ionizing source.

\subsection{Energy budget of the absorbers}
\label{energetics}

The distance constraints obtained in Section~\ref{location} allow us to estimate the amount of mass per unit time carried by the X-ray outflows in NGC 985. Assuming that the outflow moves away radially with constant speed $v$, in the form of a partial thin spherical shell at a distance $R$ from the central source, the amout of mass per unit time carried in an outflow is given by:

\begin{equation}
\label{mout}
\dot{M}_{\rm out} = \mu m_{\rm p}N_{\rm H}vR\Omega,
\end{equation}

\noindent where $\mu = 1.4$ is the mean atomic mass per proton, $m_{\rm p}$ is the proton mass, and $\Omega$ is the solid angle subtended by the outflow as seen from the central source, which ranges by definition between $0$ and $4\pi$~sr. This parameter is usually unknown, as it depends on the actual geometry of the outflow although it is typically estimated to be $\pi/2$~sr, based on the observed type-1 to type-2 ratio in nearby Seyfert galaxies (\citealt{MR95}), plus the fact that warm absorbers are detected approximately in half the observed Seyfert 1 galaxies (\citealt{Dunn07}). Indeed, this estimation for $\Omega$ is based on statistics that might be biased by selection effects (e.g., the fact that no warm absorber is detected in a given Seyfert galaxy might be due to the outflow geometry, but also to the lack of instrumental sensitivity).

From Eq.~\ref{mout} it follows that the faster, more massive winds, will carry more mass into the ISM of the host galaxy. Similarly, the outflows located further away will carry more mass per unit time than those located much closer to the central source even if the latter are denser. This is clearly exemplified in Component $A$ which has large column, fast speed, outflowing at pc-scale distances. Taking the upper limit on the distance for this component we obtain that this outflow has, at most, a mass rate of $\sim 6.5\Omega$~M$_{\odot}$~yr$^{-1}$. This is the bulk of the mass outflow rate in this AGN, several times higher than those of components $B$, $C$, and $D$, with $\dot{M}_{\rm out} \sim 0.5\Omega$, $\sim 1.2\Omega$, and $\sim 0.2\Omega$~M$_{\odot}$~yr$^{-1}$, respectively. Even the obscurer, which has comparable column density and velocity, has a considerably less impact on its surroundings because it is significantly closer to the central SMBH, with an estimated $\dot{M}_{\rm out}$ of around $\sim 0.04\Omega$~M$_{\odot}$~yr$^{-1}$, using as upper limit on its distance the 16 light-days at which the BLR is likely located based on the UV data.

The values of $\dot{M}_{\rm out}$~listed in Table~\ref{enertab} were obtained assuming that the solid angle subtended by the outflows is $\Omega = \pi/2$~sr, and must be considered upper limits, as they were calculated with the estimated upper limits on their distance. The total mass rate conveyed by the X-ray outflows is $\sim 13$~M$_{\odot}$~yr$^{-1}$, also an upper limit. For comparison, the SMBH in NGC 985 is accreting matter at a rate given by:

\begin{equation}
\label{macc}
\dot{M}_{\rm acc} = \frac{L_{\rm bol}}{\eta c^2},
\end{equation}

\noindent where $\eta = 0.1$ is the nominal accretion efficiency, $c$ is the speed of light, and $L_{\rm bol}$ is the bolometric luminosity of NGC 985. Integrating the unobscured SED we obtain $L_{\rm bol} \sim 10^{46}$~erg s$^{-1}$, which translates into a mass accretion rate of $\dot{M}_{\rm acc} = 1.76$~M$_{\odot}$~yr$^{-1}$. Therefore, the AGN in NGC 985 expels from its surroundings up to 7 times more matter than it actually accretes. This is a value often seen in Seyfert galaxies (\citealt{Blu05}; \citealt{Ebr10, Ebr16b}; see also \citealt{Cos10}), but still an order of magnitude less than other extreme cases (\citealt{Cre12}; \citealt{Ebr13}).

From the mass outflow rate is then trivial to obtain the kinetic luminosity of the outflows:

\begin{equation}
\label{klum}
L_{\rm KE} = \frac{1}{2}\dot{M}_{\rm out}v^2.
\end{equation}

\noindent It can be seen that the WA components in NGC 985 have $L_{\rm KE}$ values of $\sim 10^{41}$~erg s$^{-1}$, with the exception of component $A$, which is two orders of magnitude higher and is therefore the major contributor to the feedback in this AGN. Indeed, the kinetic luminosity of this WA component accounts for $\sim 0.8\%$ of the bolometric luminosity of the source. This would mean that this outflow by itself might be carrying enough mass and energy to significantly disrupt the ISM of NGC 985, as it is within the range $0.5\% - 5\%$ of the bolometric luminosity that is required by the models to be fed back into the medium to reproduce the observed $M - \sigma$~relation (see \citealt{HE10}, and references therein). This would then be one of the few cases in which a regular, moderate velocity warm absorber wind is deemed as a source of feedback.

It is to be emphasized that the presence of obscuring winds arising from very close to the accretion disk is possibly a more common phenomenon than originally thought as monitoring campaigns increase. This is important not only for the overall understanding of these transient events, but also for the effects on the photoionized gas that surrounds the central engine as the ionizing continuum dims and brightens again. Catching an AGN either entering or exiting an obscuration event is thus crucial to constrain the physical characteristics of the ionized gas around the central SMBH.

\begin{table}
  \centering
  \caption[]{Energetics of the X-ray WA components. The solid angle subtended by the WA is assumed to be $\Omega = \pi/2$~sr.
}
  \label{enertab}
  \begin{tabular}{l c c c c}
    \hline\hline
    \noalign{\smallskip}
    Component      &  $\dot{M}_{\rm out}$\tablefootmark{a} & $\dot{M}_{\rm out}/\dot{M}_{\rm acc}$ & $\log L_{\rm KE}$\tablefootmark{b} & $L_{\rm KE}/L_{\rm bol}$ \\
    \noalign{\smallskip}
    \hline
    \noalign{\smallskip}
    $O$  &  $0.07$  & $0.04$  & $41.9$   & $0.008$\%  \\
    $A$  &  $10.3$  & $6.5$  & $43.9$   & $0.8$\% \\
    $B$  &  $0.9$  & $0.5$  & $41.1$   & $0.001$\%  \\
    $C$  &  $1.9$  & $1.1$  & $40.9$   & $0.0008$\%   \\
    $D$  &  $0.3$  & $0.2$  & $40.5$  & $0.0003$\%  \\
    \noalign{\smallskip}
    \hline
    \noalign{\smallskip}
    Total & $13.47$ & $8.34$ & $43.93$ & $0.81$\% \\
    \noalign{\smallskip}
    \hline
  \end{tabular}
  \tablefoot{
    \tablefoottext{a}{Mass outflow rate, in units of $\rm M_{\odot}$~yr$^{-1}$; }
    \tablefoottext{b}{kinetic luminosity, in units of~erg s$^{-1}$.}
  }
\end{table}


\section{Conclusions}
\label{conclusions}

NGC 985 was observed twice by \XMM{}~in 2015 revealing that the source was coming out from a soft X-ray obscuration episode that took place in 2013 (\citealt{Par14}; \citealt{Ebr16}). In this work we have analyzed the high-resolution X-ray spectra of these observations obtained by RGS together with another \XMM{}~archival observation in 2003, when the source was mildly obscured, possibly arising from another past obscuration event.

The complex absorption spectra of NGC 985 reveals the presence of four warm absorber (WA) outflows, with column densities ranging from $\sim 10^{21}$ to a few times $10^{22}$~cm$^{-2}$. The highest ionization component $A$, with $\log \xi \simeq 2.9$, is also the fastest ($v_{\rm out} \sim -5\,100$~km s$^{-1}$). Components $B$~and $D$, with $\log \xi$ around 2.6 and 1.6, respectively, have similar kinematics ($\sim -600$~km s$^{-1}$), while component $C$ has $\log \xi = 2.0$ and is the slowest, outflowing at $\sim -350$~km s$^{-1}$. The obscurer is characterized by a mildly ionized wind ($\log \xi \sim 0.2 - 0.5$) with $N_{\rm H} \sim 2 \times 10^{22}$~cm$^{-2}$, moving away at a projected velocity of $\sim -6\,000$~km s$^{-1}$.

The multi-epoch analysis of these WA components and the obscurer showed that their parameters changed throughout the different observations, when the ionizing flux varied as the obscuration event was taking place and then progressively faded away. Assuming that the gas is in equilibrium and that the changes are due to photoionization and recombination processes in response to changes in the ionizing flux, we estimated upper and lower limits on the gas density based on the various variability and nonvariability time scales probed by the observations. These limits can then be used in turn to put stringent constraints on the location of the WA. These outflows are located at pc-scale distances (components $A$~and $B$), compatible with the inner edge of the torus, or tens of pc (components $C$~and $D$), similar to the NLR distance estimated in other AGN, from the central ionizing source. There is, however, no evidence of a connection between the emitting gas and the WA. The emission lines detected in the RGS spectra of NGC 985 are typically narrow and do not show any significant blueshift in their centroids. On the other hand, the obscurer is likely located much closer to the accretion disk, possibly close to the BLR based on the UV data.

In spite of the similar kinematics, columns and co-location on the stability curves, it is very unlikely that the obscurer and WA component $A$ belong to the same long-lived structure. It cannot be ruled out that component $A$ is actually an ejecta from the obscurer that has travelled further away until it has intersected our line of sight in the course of these observations. If we assume that it was launched close to the central engine, perhaps in the form of a disk wind, and that it has been traveling at the current measured velocity (i.e., that no other acceleration mechanism was in place), then the outflow has needed $\sim$300 years to reach the lower limit on the location of component $A$ ($\sim$2 pc). Considering all these assumptions and the long time scales involved it is unclear that there is a causal connection between these two components. On the other hand, the different limits derived on the distance for the WA, and their respective ionization parameters, may suggest that the WA is stratified, with the different components being launched from different parts of the disk.

In terms of mass and energy deposited back into the ISM of NGC 985, WA component $A$ carries the bulk of the mass and the kinetic energy per unit time, owing to its distance to the AGN engine, density, and velocity. This component by itself could carry enough momentum to disturb the ISM, as its kinetic luminosity is about $0.8\%$ of the bolometric luminosity of the AGN. This value is within the range of $L_{\rm KE}/L_{\rm bol}$~fraction required by many models to account for cosmic feedback.

Observations of the onset and the ending of transient eclipsing events in AGN is therefore a key opportunity to study the obscuring matter itself, but also the impact on the surrounding gas as the ionizing flux varies.


\begin{acknowledgements}
This work is based on observations obtained by \XMM{}, an ESA science mission with instruments and contributions directly funded by ESA member states and the USA (NASA). V.D. acknowledges support from the ESA Trainee program and the ESAC Space Science Faculty. This work was supported by NASA through a grant for HST program number 13812 from the Space Telescope Science Institute, which is operated by the Association of Universities for Research in Astronomy, Incorporated, under NASA contract NAS5-26555. SRON is supported financially by NWO, the Netherlands Organization for Scientific Research.
\end{acknowledgements}

%
%


\bibliographystyle{aa}
\bibliography{references}



\begin{appendix}
\section{\XMM{}~RGS spectra of NGC 985}
\label{rgsspectra}
  
\begin{figure*}[t]
  \centering
  \hbox{
  \includegraphics[width=6.5cm,angle=-90]{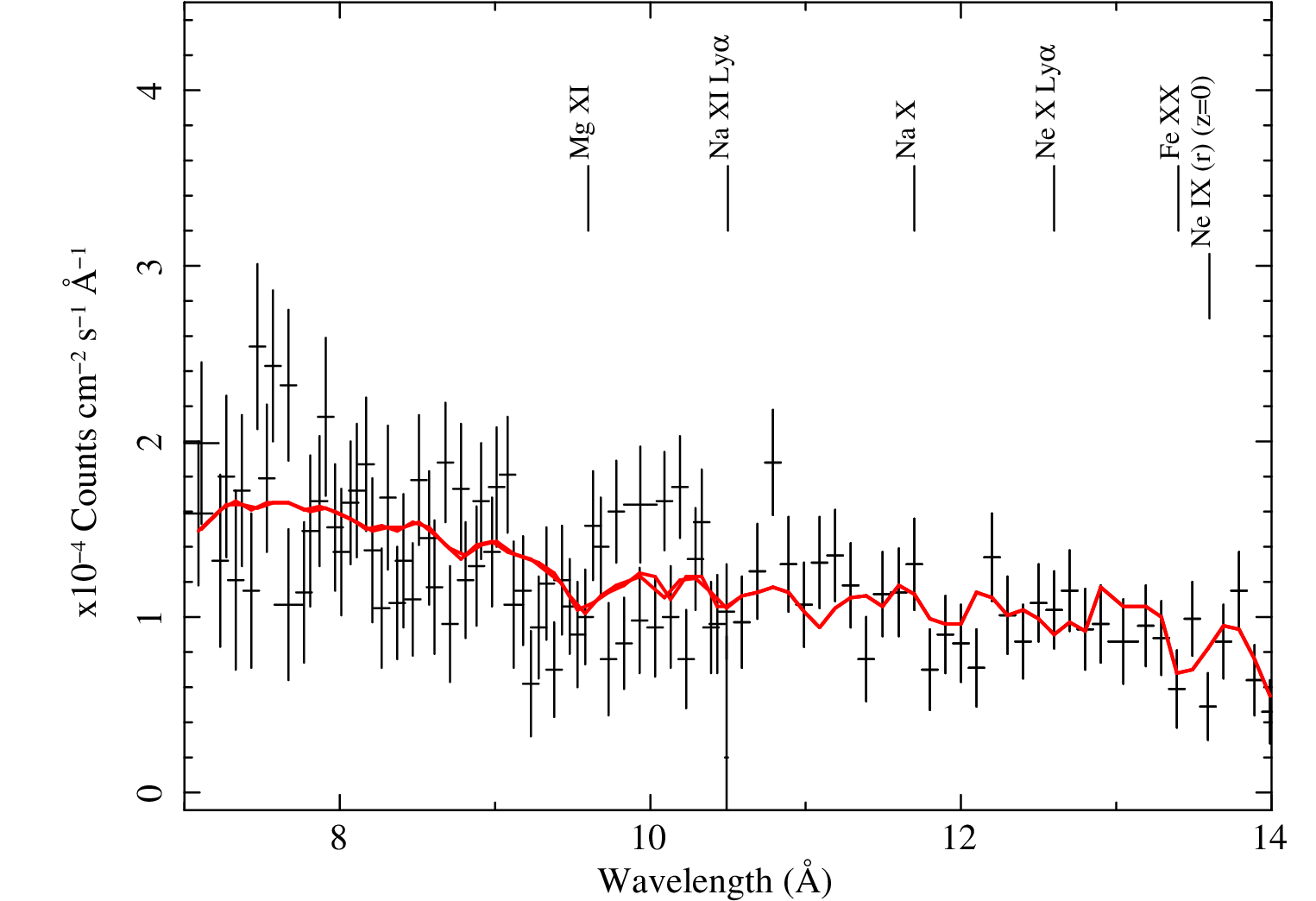}
  \includegraphics[width=6.5cm,angle=-90]{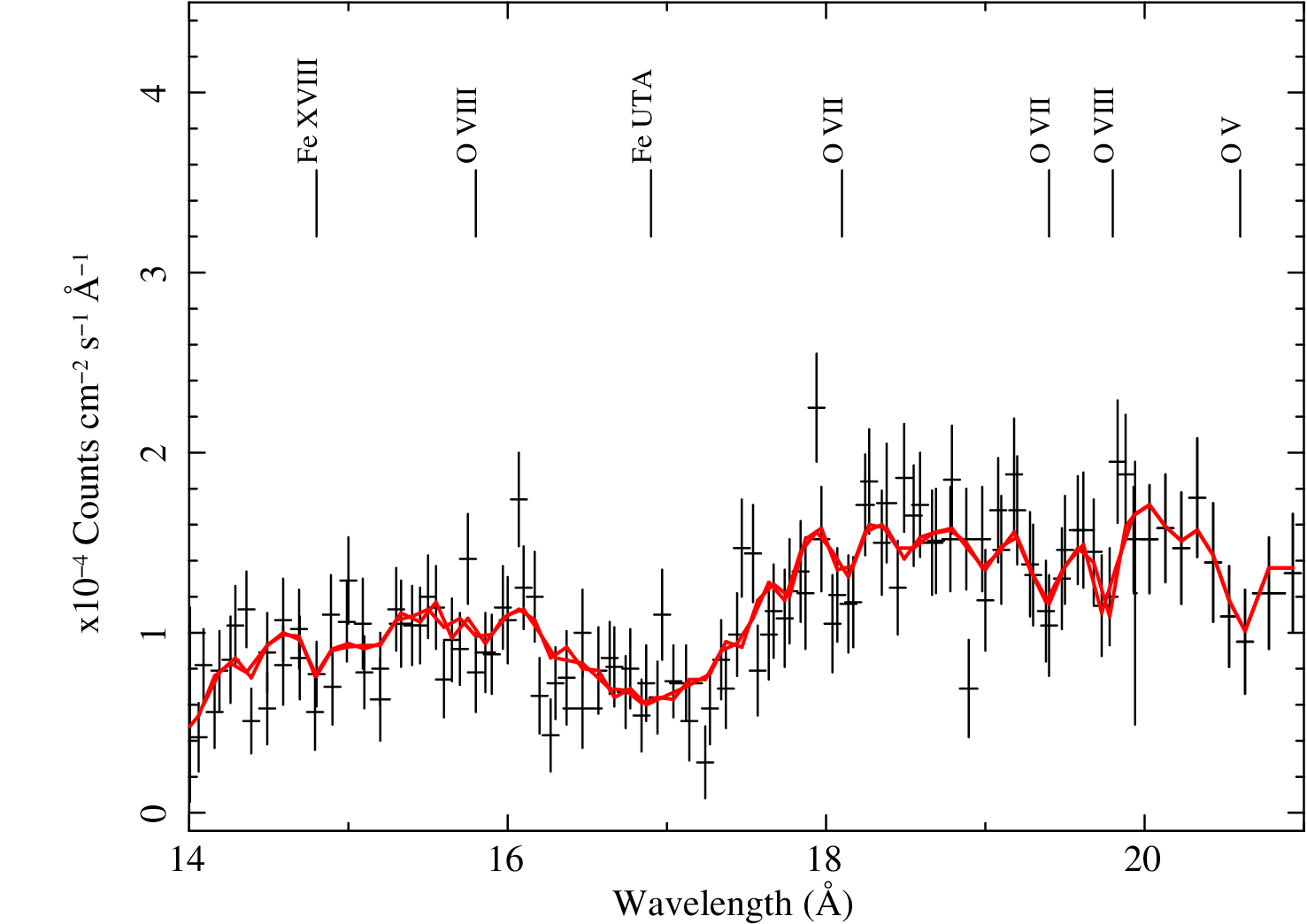}
  }
  \hbox{
  \includegraphics[width=6.5cm,angle=-90]{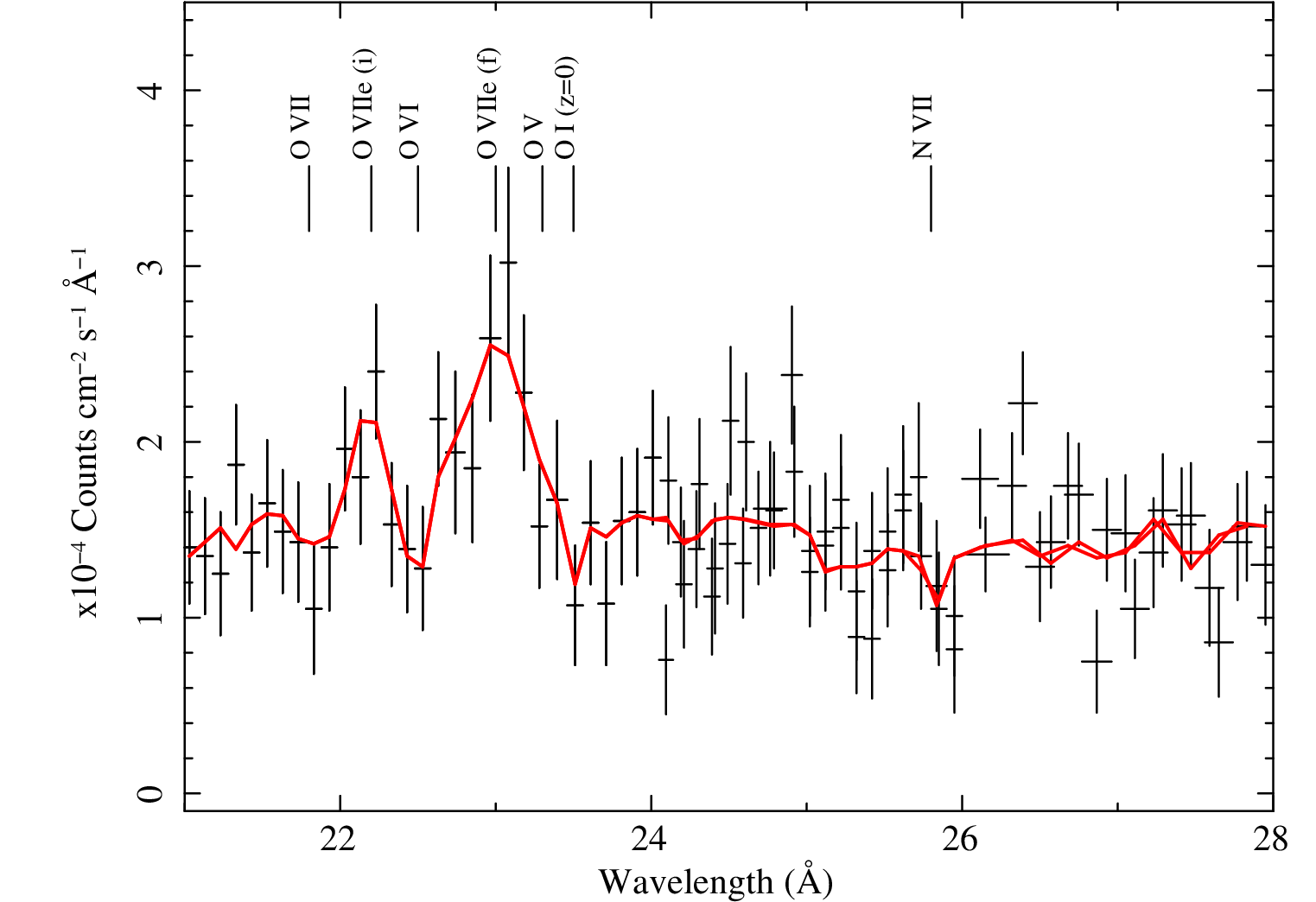}
  \includegraphics[width=6.5cm,angle=-90]{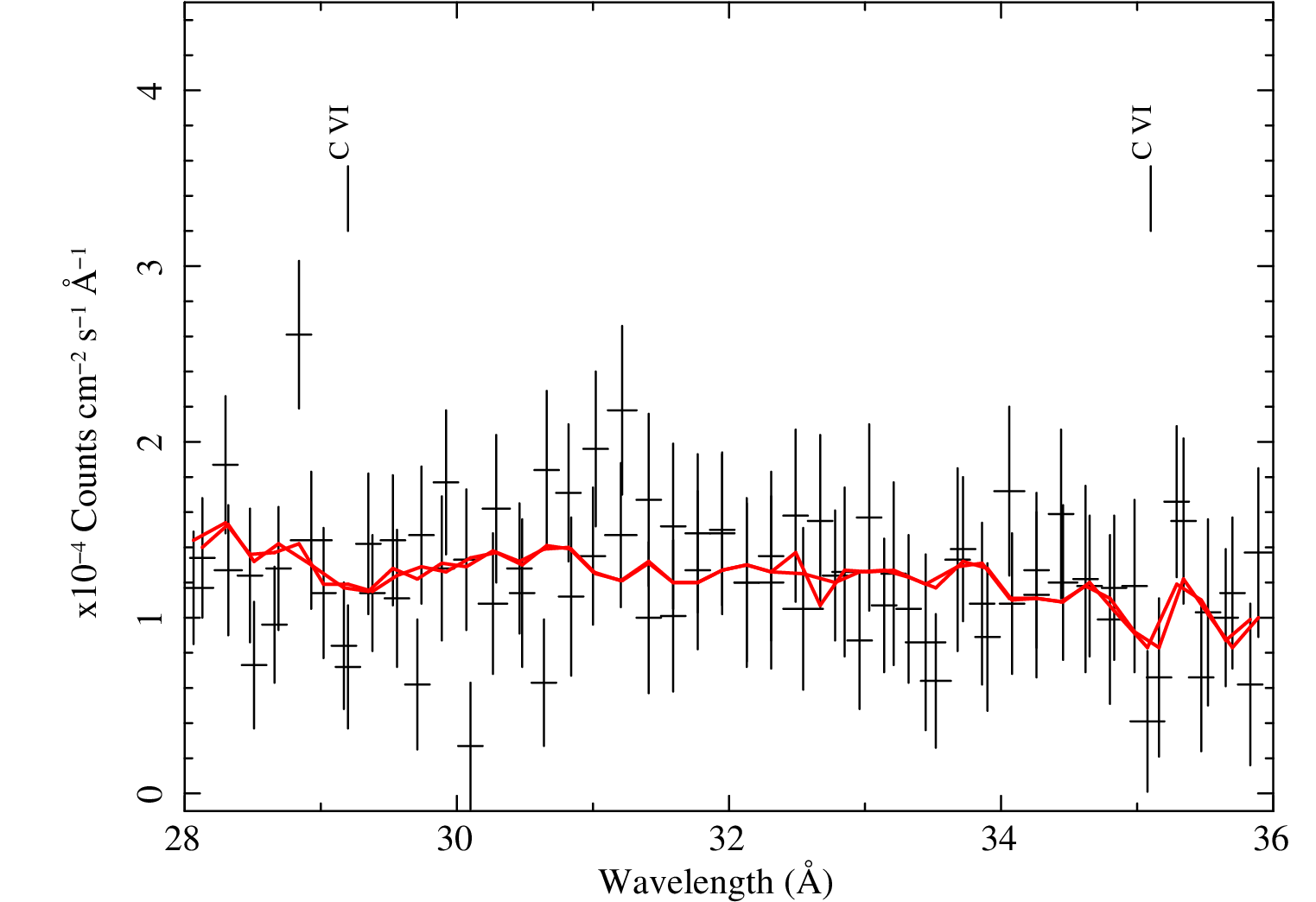}
  }
  \caption{\label{spec03}\XMM{}~RGS spectrum of NGC 985 in 2003. The solid red line represents the best fit model. The most relevant features have been labeled.}
\end{figure*}

\begin{figure*}[t]
  \centering
  \hbox{
  \includegraphics[width=6.5cm,angle=-90]{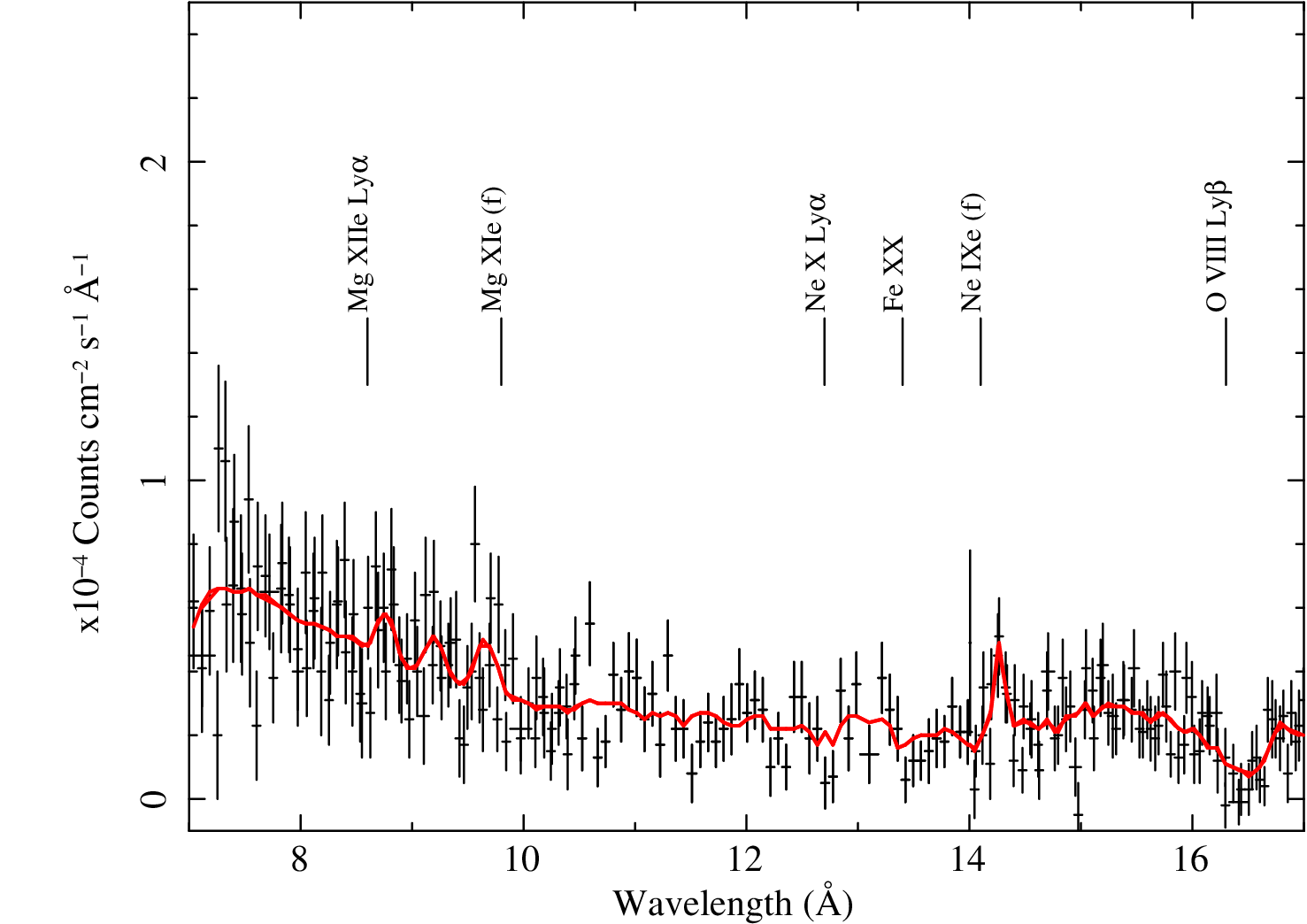}
  \includegraphics[width=6.5cm,angle=-90]{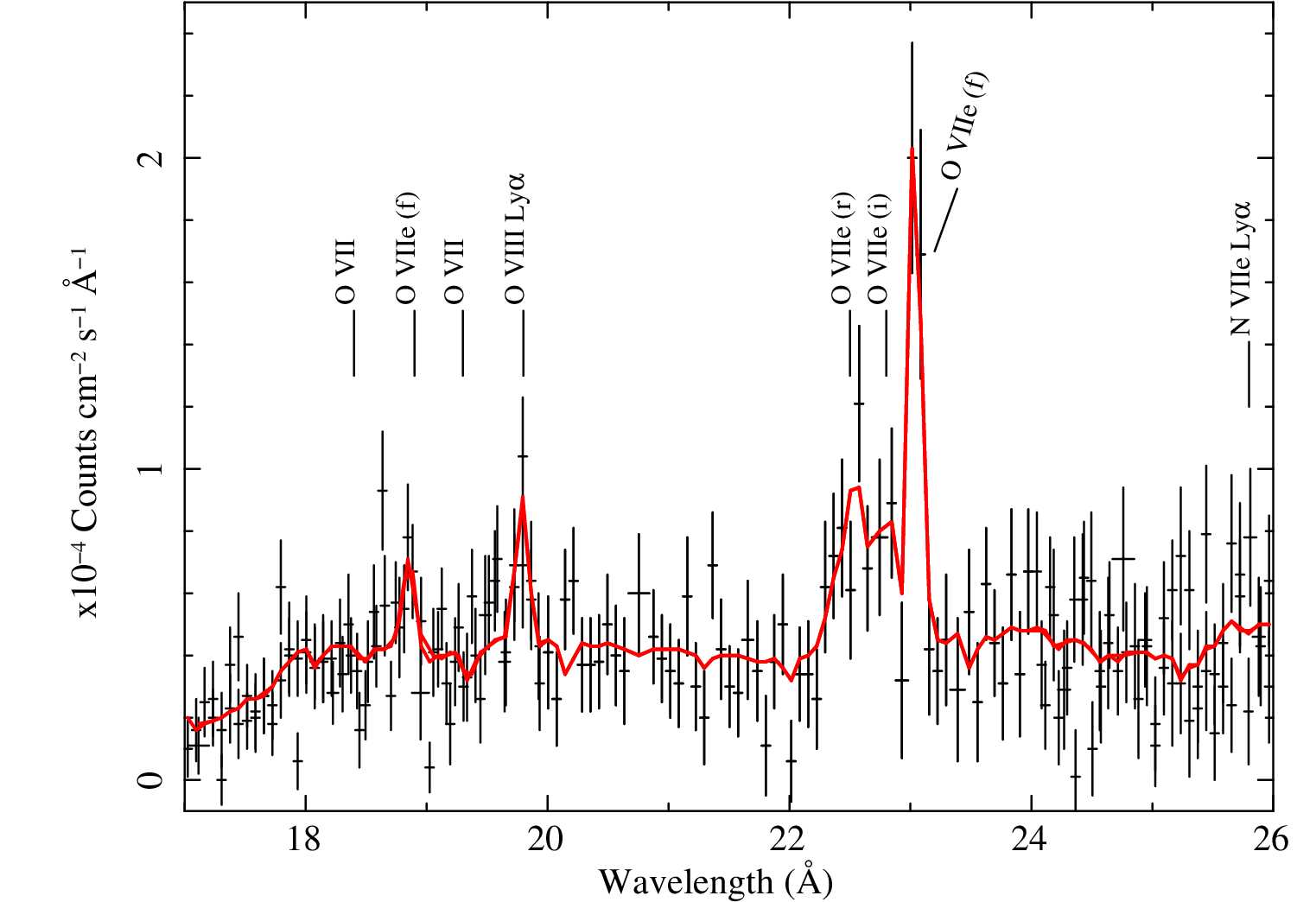}
  }
  \caption{\label{spec13}\XMM{}~RGS spectrum of NGC 985 in 2013. The solid red line represents the best fit model. The spectrum has been rebinned for clarity. The most relevant features have been labeled.}
\end{figure*}

\begin{figure*}[t]
  \centering
  \hbox{
  \includegraphics[width=6.5cm,angle=-90]{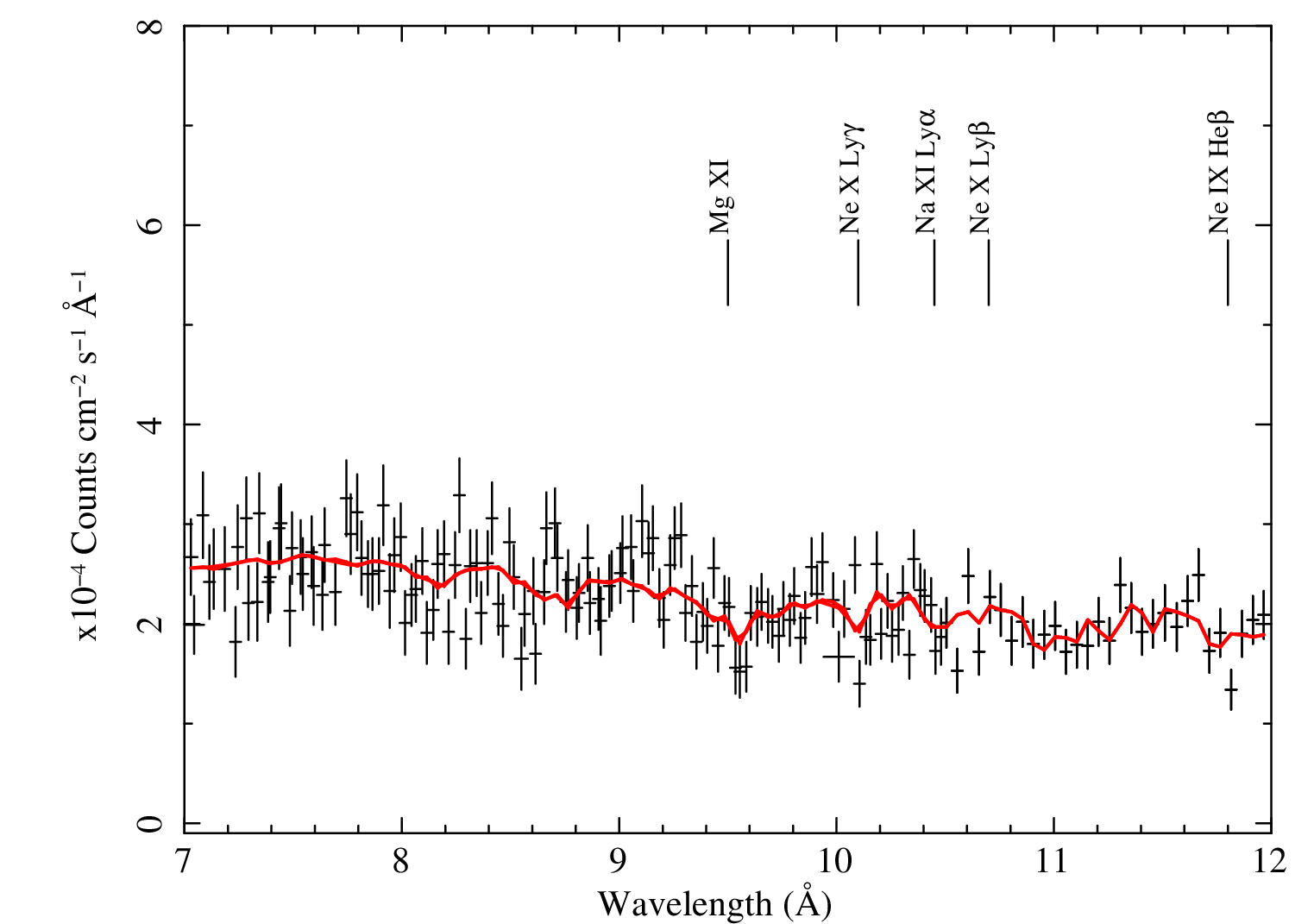}
  \includegraphics[width=6.5cm,angle=-90]{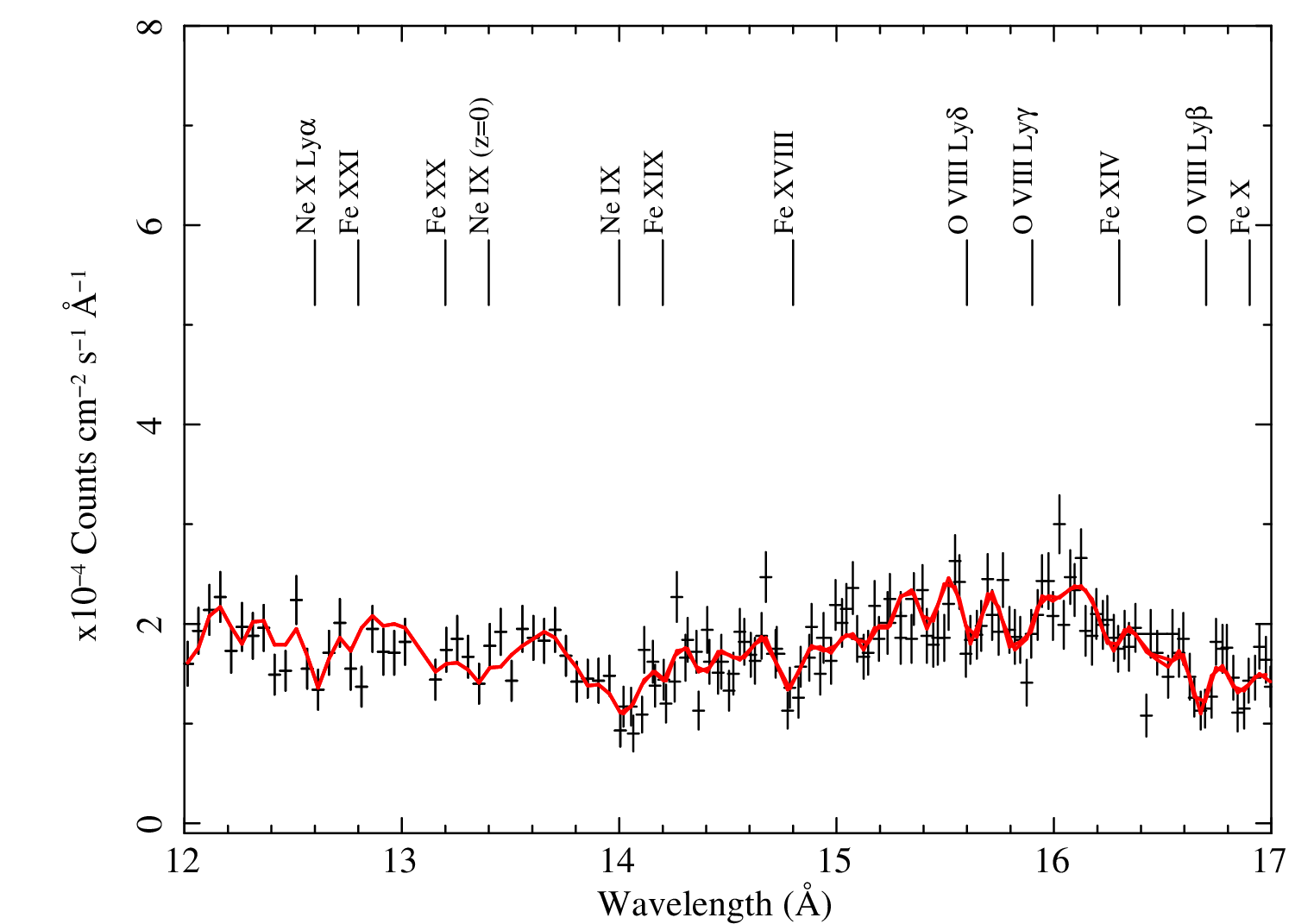}
  }
  \hbox{
  \includegraphics[width=6.5cm,angle=-90]{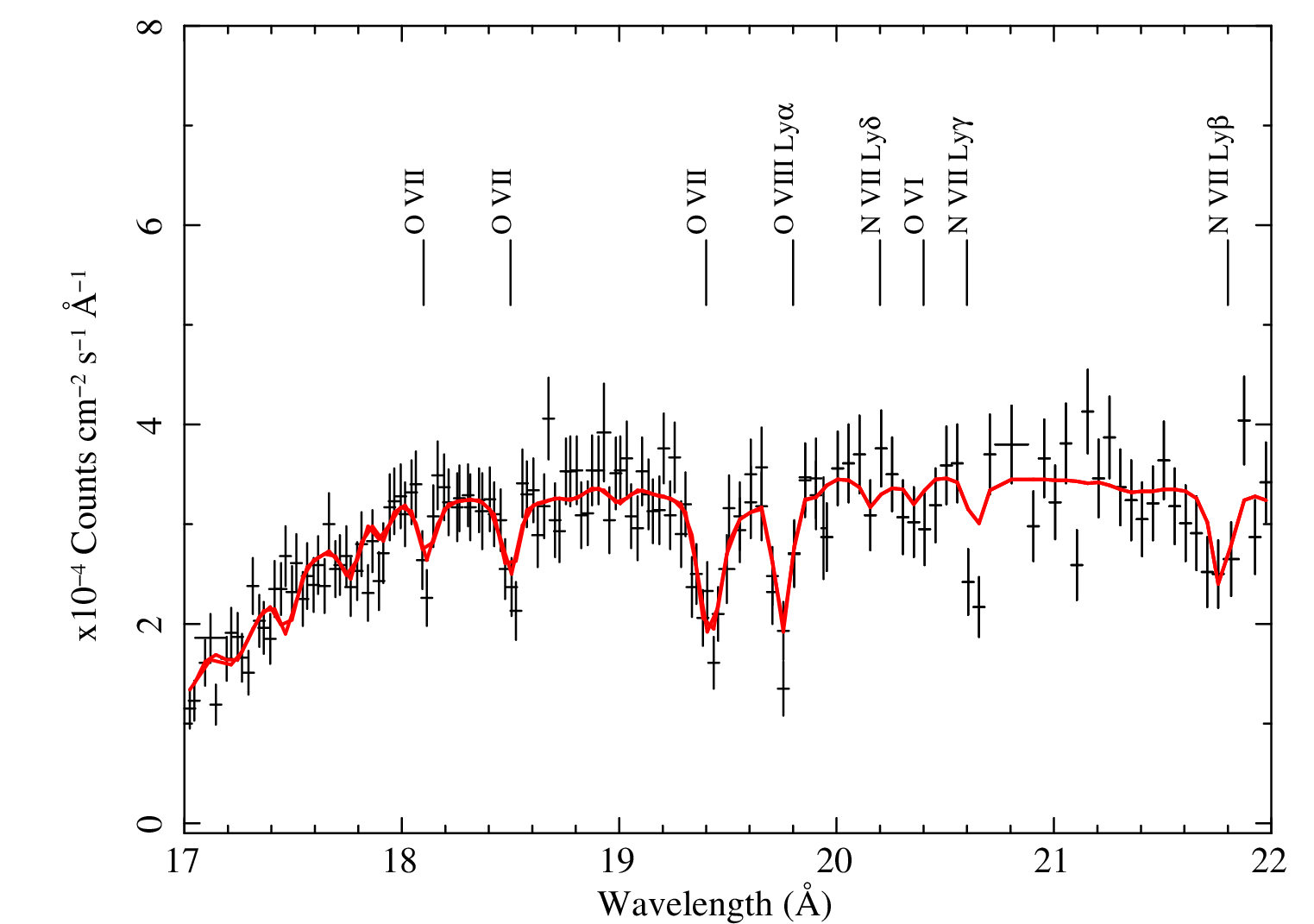}
  \includegraphics[width=6.5cm,angle=-90]{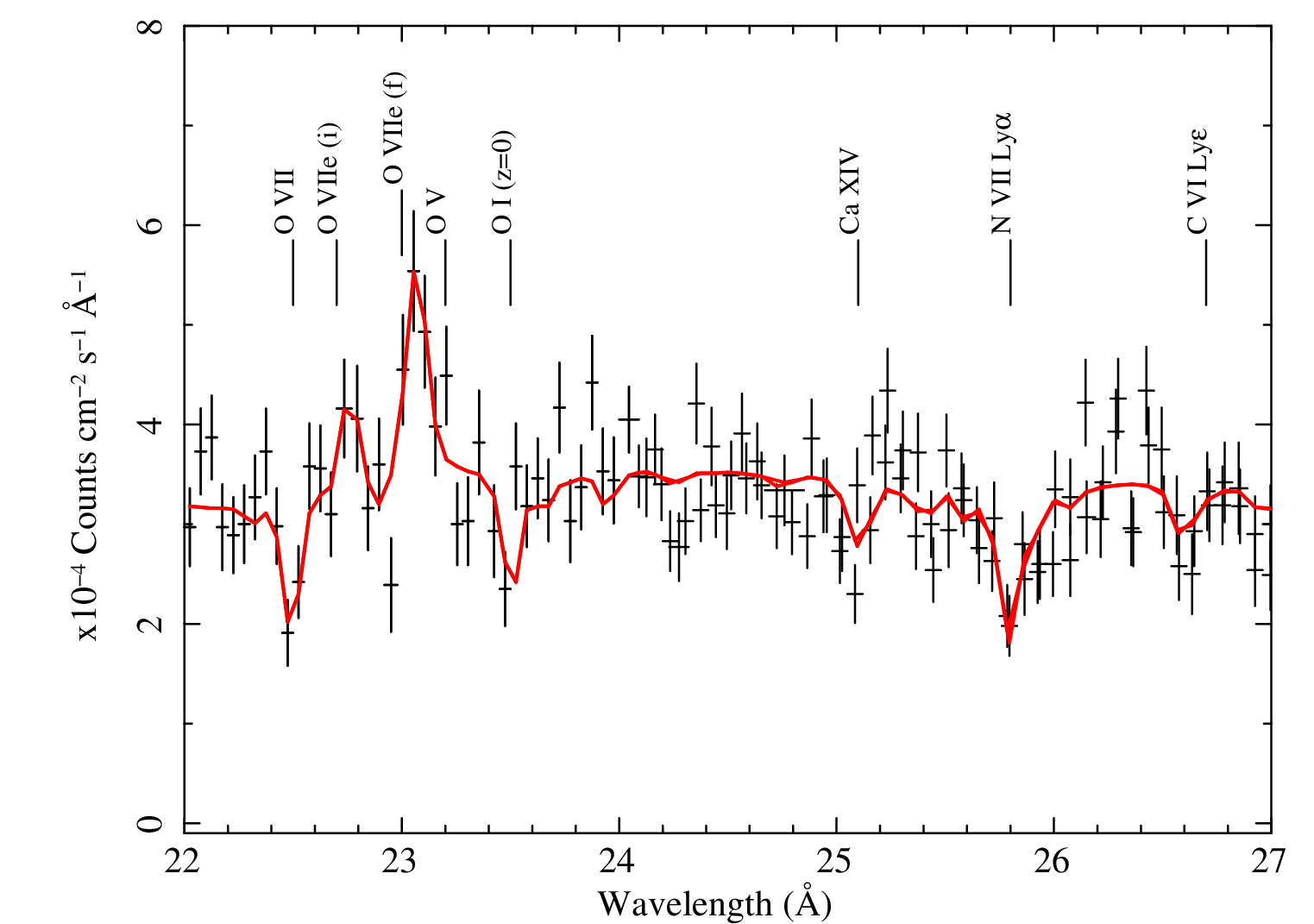}
  }
  \hbox{
  \includegraphics[width=6.5cm,angle=-90]{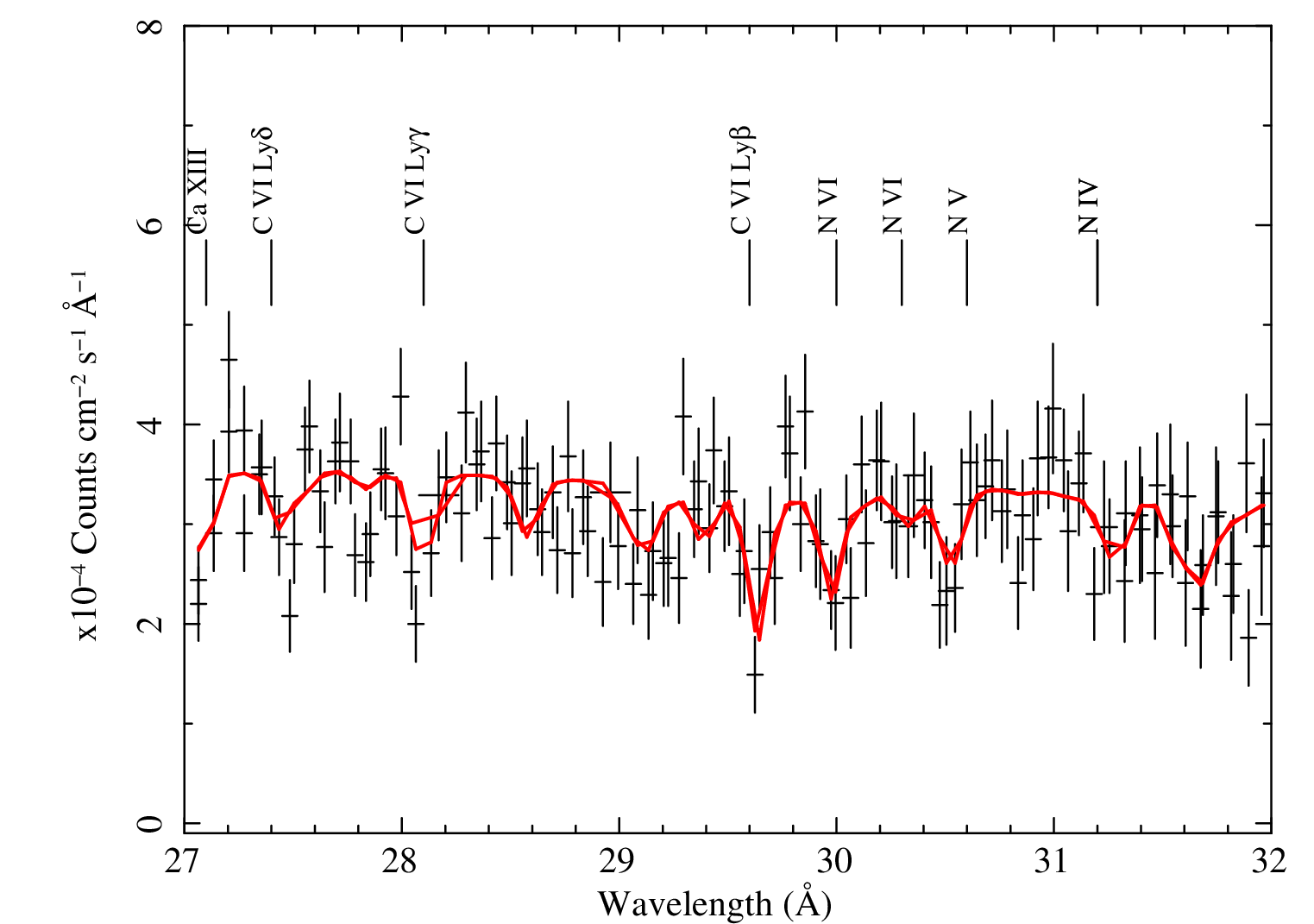}
  \includegraphics[width=6.5cm,angle=-90]{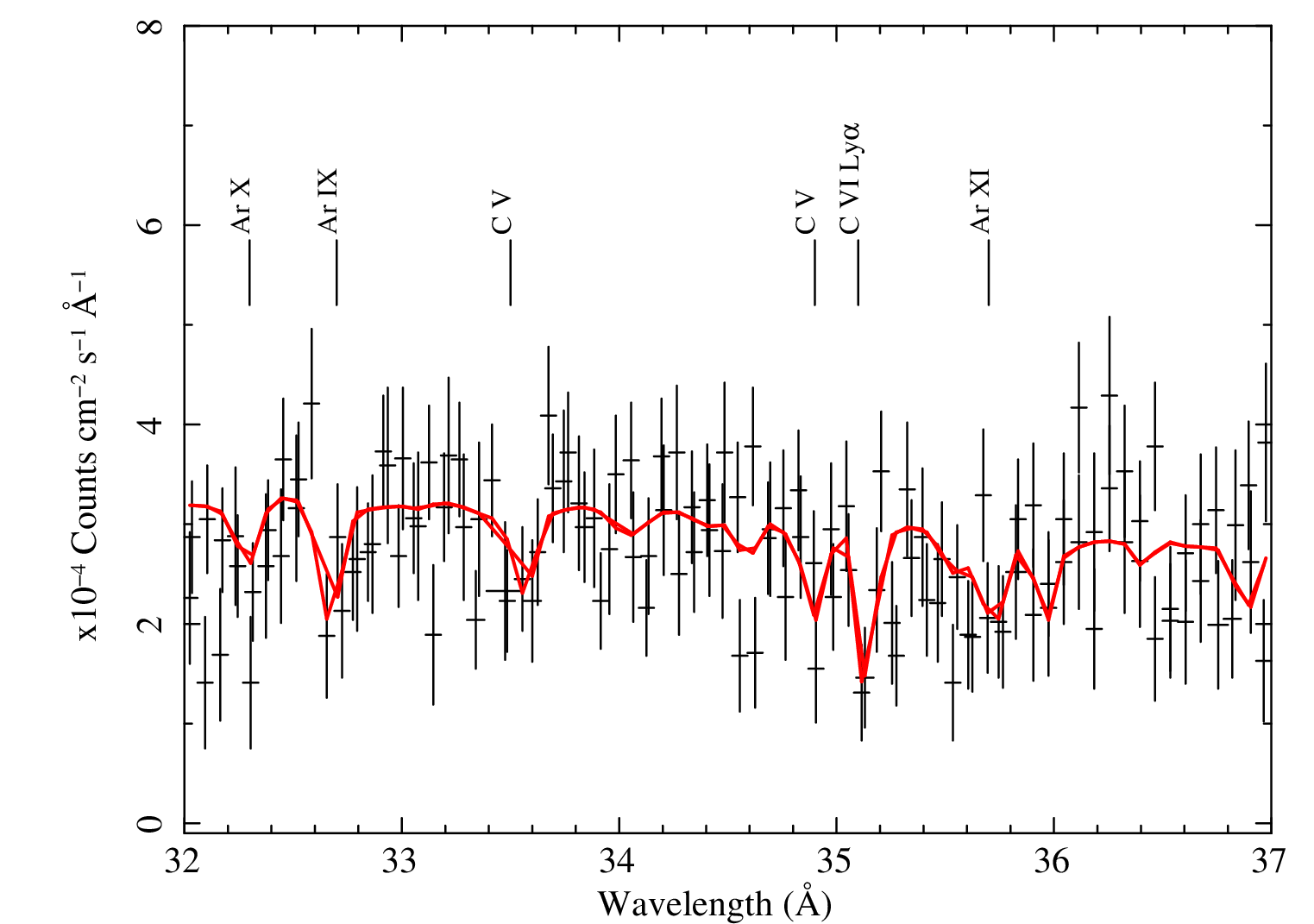}
  }
  \caption{\label{spec15a}\XMM{}~RGS spectrum of NGC 985 in 2015 (first observation). The solid red line represents the best fit model. The most relevant features have been labeled.}
\end{figure*}

In this Appendix section we show the RGS spectra used in this work together with their best-fit model. The 2003 observation is shown in Fig.~\ref{spec03}, the 2013 observation in Fig.~\ref{spec13}, and the first observation of 2015 in Fig.~\ref{spec15a}.

\section{Plots of the absorption models}
\label{modelplots}

In this Appendix section we show the model plots of the different absorption comoponents detected in the \XMM{} observations. In this way we show the contribution in different parts of the spectra of the different components. The different absorption components have been shifted in flux for clarity.

\begin{figure*}[t]
  \centering
  \hbox{
    \includegraphics[width=6.5cm,angle=-90]{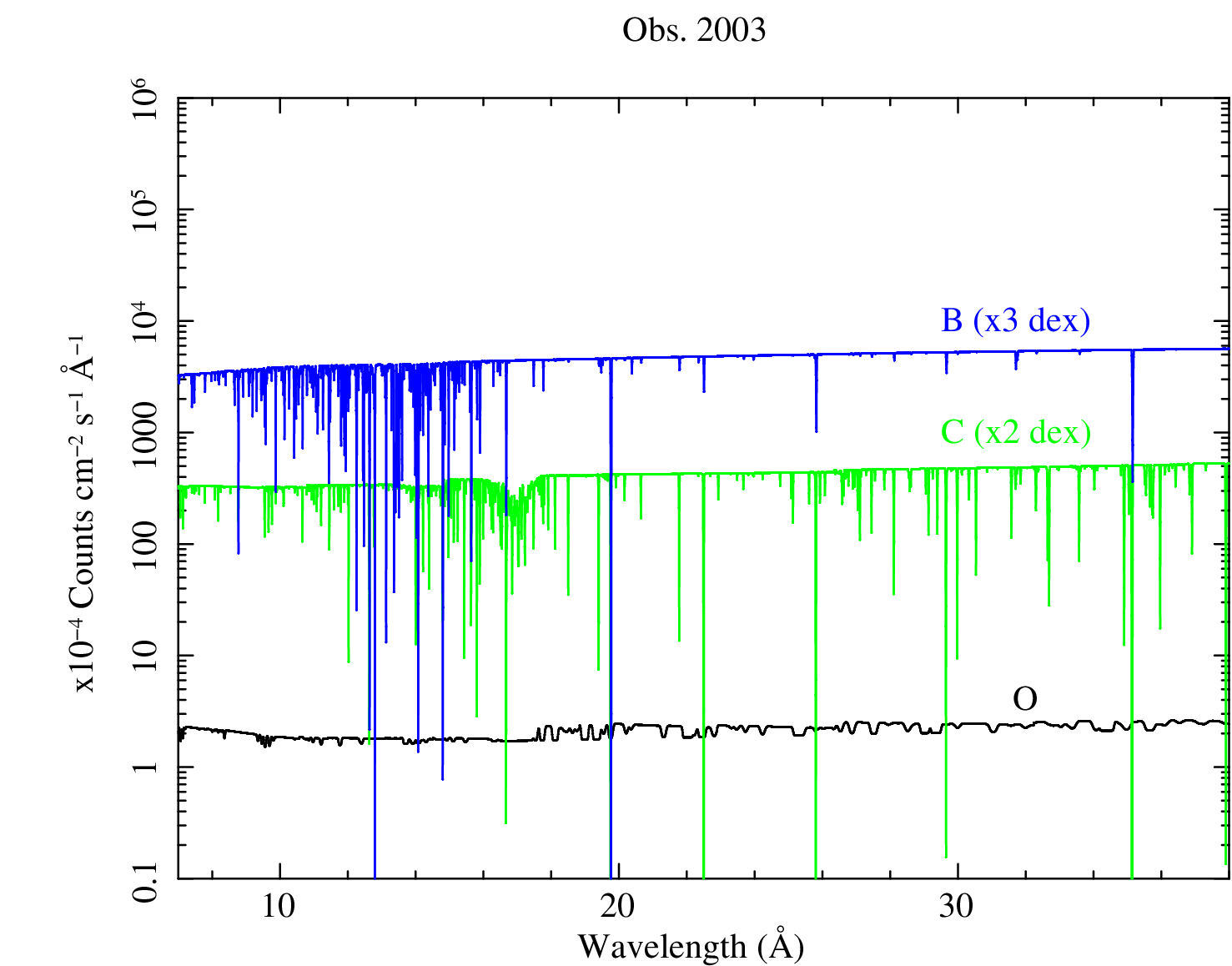}
    \includegraphics[width=6.5cm,angle=-90]{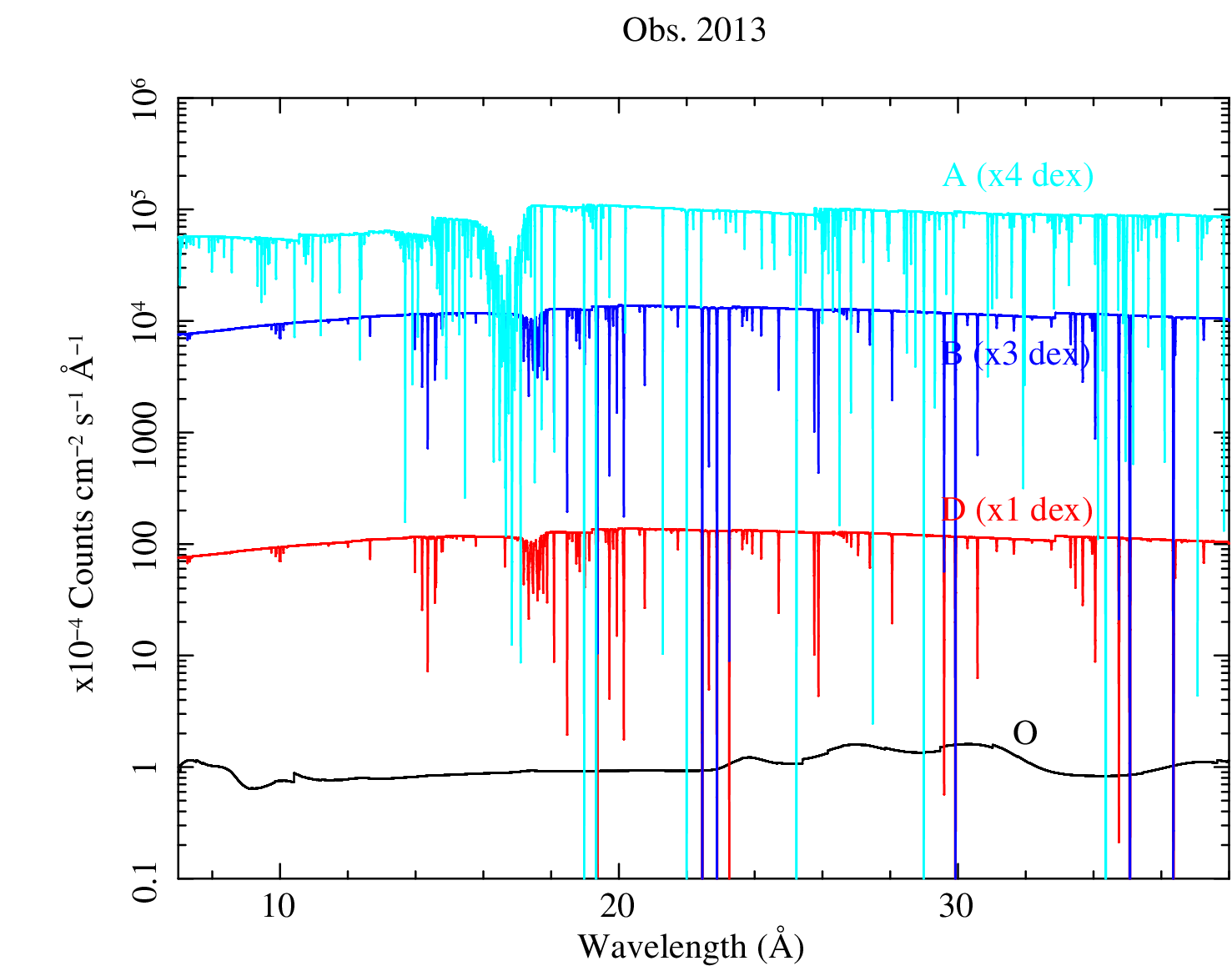}
  }
  \hbox{
    \includegraphics[width=6.5cm,angle=-90]{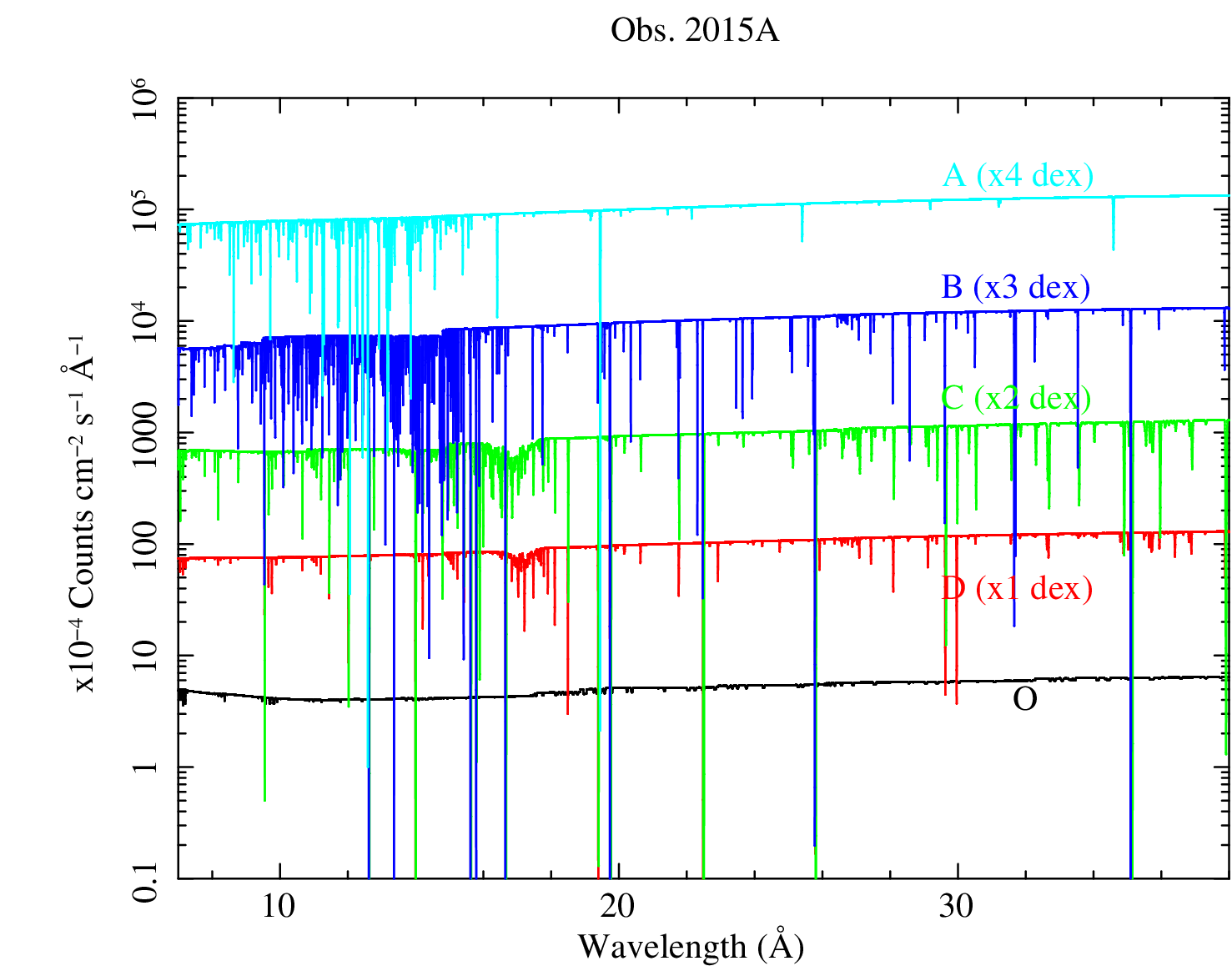}
    \includegraphics[width=6.5cm,angle=-90]{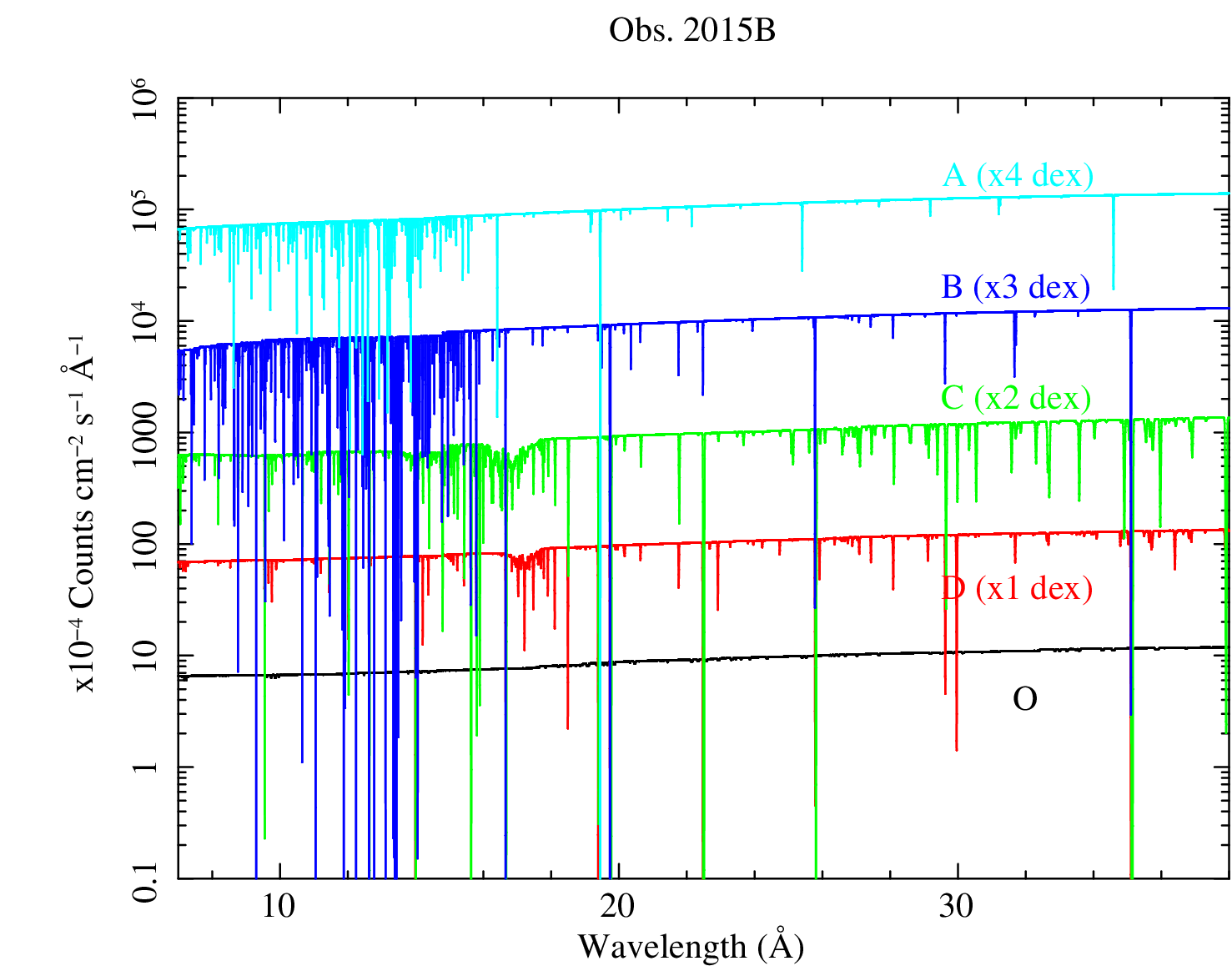}
  }
  \caption{\label{absmodels}{Model plot showing the contribution to the RGS spectrum of each of the {\it xabs} components that model the absorption troughs seen in NGC 985 for the observations in 2003 ({\it upper left panel}), 2013 ({\it upper right panel}), 2015a ({\it lower left panel}), and 2015b ({\it lower right panel}). For clarity, the contribution of Component $D$ has been shifted by 1 dex in flux, the one of Component $C$ by 2 dex, the one of Component $B$ by 3 dex, and the one of Component $A$ by 4 dex, whenever they have been detected.}}
\end{figure*}


\section{Ionic species used for variability calculations}
\label{varions}

In Tables~\ref{ionlistA} to~\ref{ionlistD} we list the ions that were used in Sect.~\ref{variability} as a proxy for each WA component, and their corresponding product $nt_{\rm rec}$ as provided by the SPEX auxiliary program {\it rec\_time}.

\begin{table}[hb]
  \centering
  \caption[]{Ions used as a proxy for the WA component $A$.}
  \label{ionlistA}
  \begin{tabular}{l c}
    \hline\hline
    \noalign{\smallskip}
    Ion   & $nt_{\rm rec}$ (s\,cm$^{-3}$)\\
    \noalign{\smallskip}
    \hline
    \noalign{\smallskip}
    \ion{Na}{xi} &  $7.6 \times 10^{10}$ \\
    \ion{Mg}{xii} & $1.2 \times 10^{11}$ \\
    \ion{Fe}{xx} & $3.1 \times 10^{10}$ \\
    \ion{Fe}{xxiv} & $5.4 \times 10^{10}$ \\
    \noalign{\smallskip}
    \hline
  \end{tabular}
\end{table}

\begin{table}[hb]
  \centering
  \caption[]{Ions used as a proxy for the WA component $B$.}
  \label{ionlistB}
  \begin{tabular}{l c}
    \hline\hline
    \noalign{\smallskip}
    Ion   &   $nt_{\rm rec}$ (s\,cm$^{-3}$)  \\
    \noalign{\smallskip}
    \hline
    \noalign{\smallskip}
    \ion{N}{vii} &  $1.6 \times 10^{10}$ \\
    \ion{O}{viii} &  $3.0 \times 10^{10}$ \\
    \ion{Ne}{x} &  $8.8 \times 10^{10}$ \\
    \ion{Na}{xi} &  $1.7 \times 10^{11}$ \\
    \ion{Mg}{xi} &  $6.2 \times 10^{10}$ \\
    \ion{Mg}{xii} &  $8.0 \times 10^{10}$ \\
    \ion{Si}{xiii} &  $1.8 \times 10^{11}$ \\
    \ion{Fe}{xviii} &  $2.3 \times 10^{10}$ \\
    \ion{Fe}{xix} &  $2.2 \times 10^{10}$ \\
    \ion{Fe}{xx} &  $1.4 \times 10^{10}$ \\
    \noalign{\smallskip}
    \hline
  \end{tabular}
\end{table}

\begin{table}[hb]
  \centering
  \caption[]{Ions used as a proxy for the WA component $C$.}
  \label{ionlistC}
  \begin{tabular}{l c}
    \hline\hline
    \noalign{\smallskip}
    Ion   &   $nt_{\rm rec}$ (s\,cm$^{-3}$)  \\
    \noalign{\smallskip}
    \hline
    \noalign{\smallskip}
    \ion{C}{vi} &  $3.0 \times 10^{10}$ \\
    \ion{N}{vi} &  $1.4 \times 10^{11}$ \\
    \ion{N}{vii} &  $7.7 \times 10^{10}$ \\
    \ion{O}{vii} &  $2.4 \times 10^{10}$ \\
    \ion{O}{viii} &  $1.1 \times 10^{11}$ \\
    \ion{Ne}{ix} &  $1.4 \times 10^{11}$ \\
    \ion{Ne}{x} &  $1.1 \times 10^{11}$ \\
    \ion{Mg}{xi} &  $5.2 \times 10^{10}$ \\
    \ion{Ar}{xi} &  $2.5 \times 10^{10}$ \\
    \ion{Ar}{xii} &  $9.6 \times 10^{10}$ \\
    \ion{Ar}{xiii} &  $3.8 \times 10^{10}$ \\
    \ion{Ca}{xiii} &  $5.0 \times 10^{10}$ \\
    \ion{Fe}{x}  & $2.8 \times 10^{9}$ \\
    \ion{Fe}{xi} &  $3.2 \times 10^{9}$ \\
    \ion{Fe}{xii} &  $2.6 \times 10^{10}$ \\
    \ion{Fe}{xiii} &  $4.7 \times 10^{9}$ \\
    \noalign{\smallskip}
    \hline
  \end{tabular}
\end{table}

\begin{table}[hb]
  \centering
  \caption[]{Ions used as a proxy for the WA component $D$.}
  \label{ionlistD}
  \begin{tabular}{l c}
    \hline\hline
    \noalign{\smallskip}
    Ion   &   $nt_{\rm rec}$ (s\,cm$^{-3}$)  \\
    \noalign{\smallskip}
    \hline
    \noalign{\smallskip}
    \ion{C}{vi} &  $1.0 \times 10^{11}$ \\
    \ion{N}{vi} &  $5.5 \times 10^{10}$ \\
    \ion{O}{vii} &  $9.5 \times 10^{10}$ \\
    \ion{Ar}{xi} &  $6.0 \times 10^{10}$ \\
    \ion{Fe}{viii} &  $2.7 \times 10^{10}$ \\
    \ion{Fe}{ix} &  $2.1 \times 10^{9}$ \\
    \ion{Fe}{x} &  $3.8 \times 10^{11}$ \\
    \noalign{\smallskip}
    \hline
  \end{tabular}
\end{table}

\end{appendix}


\end{document}